\documentclass[aps,pre,showpacs,twocolumn]{revtex4}
\usepackage{bm}
\usepackage{graphicx}
\bibstyle{approve.bib}
\usepackage{amssymb}
\usepackage{amsmath}
\usepackage{esint}
\usepackage{epstopdf}
\usepackage{color}
\usepackage{gensymb}
\usepackage{mathtools}
\usepackage{suffix}
\usepackage{fancyhdr}

\newcommand{\be}{\begin{equation}}
\newcommand{\ee}{\end{equation}}
\newcommand{\bea}{\begin{eqnarray}}
\newcommand{\eea}{\end{eqnarray}}

\newcommand{\br}{\mathbf{r}}

\newcommand{\bE}{\mathbf{E}}

\newcommand{\e}{\varepsilon}

\newcommand{\tG}{\tilde{G}}

\newcommand{\mI}{\mathrm{I}_m}
\newcommand{\mK}{\mathrm{K}_m}

\newcommand{\m}{_{\rm m}}
\newcommand{\ce}{_{\rm c}}

\newcommand{\s}{_{\rm s}}

\newcommand{\hu}{\hat{u}}

\newcommand{\B}{_{\rm b}}

\newcommand{\SB}[1]{\textcolor{black} {#1}}
\newcommand{\BS}[1]{\textcolor{black} {#1}}

\begin{document}

\title{Dielectric manipulation of polymer translocation dynamics in \BS{engineered membrane nanopores}}

\author{Sahin Buyukdagli}
\address{Department of Physics, Bilkent University, Ankara 06800, Turkey}

\begin{abstract}

\noindent{\bf \SB{ABSTRACT:}} The alteration of the dielectric membrane properties by \BS{membrane engineering techniques such as carbon nanotube (CNT) coating} opens the way to novel molecular transport strategies for biosensing purposes. In this article, we predict a macromolecular transport mechanism enabling the dielectric manipulation of the polymer translocation dynamics in \BS{dielectric membrane} pores confining mixed electrolytes.  In the giant permittivity regime of \BS{these engineered} membranes governed by attractive polarization forces, multivalent ions adsorbed by the membrane nanopore trigger a monovalent ion separation and set an electroosmotic \BS{counterion} flow. The drag force exerted by this flow is sufficiently strong to suppress and invert the electrophoretic velocity of anionic polymers, and also to generate the mobility of neutral polymers whose speed and direction can be solely adjusted by the charge and concentration of the added multivalent ions. These features identify the dielectrically generated transport mechanism as an efficient mean to drive overall neutral or weakly charged analytes that cannot be controlled by an external voltage. We also reveal that in anionic polymer translocation, multivalent cation addition into the monovalent salt solution amplifies the electric current signal by several factors. The signal amplification is caused by the electrostatic many-body interactions replacing the monovalent polymer counterions by the multivalent cations of higher electric mobility. The strength of this elektrokinetic charge discrimination points out the potential of multivalent ions as current amplifiers capable of providing boosted resolution in nanopore-based biosensing techniques.
\end{abstract}

\date{\today}
\maketitle   

\section{Introduction}

Macromolecular transport under nanoscale confinement lies at the heart of various {\it in vivo} and {\it in vitro} biological phenomena. From the artificial delivery of genetic material into the cell medium~\cite{gn1,gn2,gn3} to viral infection~\cite{biomatter} and nanopore-based biosequencing~\cite{rev2,com}, our understanding and control of these processes rely on the accurate characterization of the macromolecular dynamics in the confined liquid environment. Among the aforementioned processes, the characterization and optimization of the field-driven DNA translocation is a crucial technical step to meet this challenge. However, due to the diversity of the competing electrostatic, hydrodynamic, and entropic effects driving the confined polymer-liquid complex, a complete prescription for the sequence detection by translocation is still missing~\cite{Tapsarev}.

The experimentally gained insight into the nanopore sensing approaches indicates that the primary requirement for the resolution of single nucleotides is a strong current signal of long duration~\cite{e1,e2,e6,e4,e5,e3,e7,e8,e9,e10,Coating1,Coating2,Coating3}. This technical constraint implies the necessity to identify the physical conditions maximizing the mean dwell time spent by the analyte inside the nanopore. Accounting for the effect of polymer conformations and steric polymer-membrane interactions on the translocation process, the connection between the dwell time and the polymer length has been initially characterized by numerical studies~\cite{n1,n2,n5}, scaling approaches~\cite{Wolterink}, and the tension propagation theory~\cite{Saka1,Saka2}. Subsequently, the consideration of the electrostatic and non-equilibrium nature of the translocation process led to modified Smoluchowski formalisms~\cite{mut1,mut2,mut3,Chinappi2,HLsim,cap1,cap2} and simulation models~\cite{aks1,aks2,aks3,Hsiao} probing the effect of the electrostatic control parameters such as the external voltage, salt, and the acidity of the solution on the polymer capture and translocation dynamics.

Field-driven polymer translocation occurs via the collective motion of the macromolecule and the surrounding salt ions hydrated in water. This implies that the quantitatively accurate characterization of polymer translocation requires the modeling of the pore electrostatics and hydrodynamics on an equal footing. Based on the coupling of the Stokes relation and the linearized mean-field (MF) Poisson-Boltzmann (PB) \SB{equation}, Ghosal developed the first electrohydrodynamic (EH) theory of polymer translocation through solid-state pores~\cite{Ghosal2006,Ghosal2007}. Within the one-loop theory of electrostatic interactions, we incorporated into this EH formalism the weak-coupling (WC) ion correlations~\cite{the16}. Finally, by taking into account the \SB{additional effect of the} electrostatic polymer-membrane interactions, we showed that the presence of multivalent ions enables faster polymer capture and slower translocation by like-charge polymer-membrane attraction~\cite{Buyuk2018,Buyuk2018II}.

In interfacially charged solid-state nanopores, the polymer velocity can be \SB{equally} altered by the action of the \SB{electroosmotic (EO)} flows~\cite{aks2}. However, the control of these flows requires the adjustment of the nanopore surface charge via the acidity of the buffer solution, which equally modifies the polymer surface charge subjected to protonation~\cite{Firnkes}. The resulting complexity of the $p$H-dependent translocation rates~\cite{Buyuk2018III} indicates the necessity to tune the translocation speed without activating the surface chemistry of the polymer. In this article, we show that via the adequate choice of the electrolyte composition, the dielectric properties of \BS{engineered substrates such as CNT-coated membranes~\cite{Coating4,Coating5}} allow to generate EO flows whose magnitude and direction can be controlled without affecting the polymer surface charge.

The use of coated-membranes for polymer translocation purposes has been initially motivated by the need to prevent the undesired adhesion of the translocating molecule to the membrane surface~\cite{Coating1}. Further studies revealed that membrane coating equally enables the modification of the threading force on DNA~\cite{Coating2} and the reduction of the translocation velocity by edge-pinning the molecule at the pore mouth~\cite{Coating3}. The coating of polymer-based matrices by CNTs has been independently initiated by the technological requirement to fabricate compact micro capacitors of high dielectric permittivity. Indeed, the dressing of the membrane surfaces with CNTs was revealed to amplify the dielectric permittivity of the substrate by orders of magnitude~\cite{Coating4,Coating5}.

In the present work, we explore the potential of \BS{dielectric} membrane nanopores confining electrolyte mixtures for polymer translocation purposes. At this point, it should be noted that the accurate modeling of mixed electrolytes requires the asymmetric treatment of the ionic valencies characterized by different electrostatic coupling strengths~\cite{NetzSC,Podgornik2010,Buyuk2019}.  With the aim to investigate correlation and polarization effects on nanofluidic ion transport through planar slits~\cite{Boc1,Boc2,Heyden2006,Gillespie2013}, we have recently developed the SC (strong coupling)-dressed PB (SCPB) theory consistently treating the mono- and multivalent ions within the WC and SC electrostatics, respectively~\cite{JPCB2020}. Here, we adapt the SCPB approach to the cylindrical geometry of the polymer-nanopore complex. \SB{By} coupling this electrostatic formalism with the hydrodynamic Stokes \SB{equation} and the force-balance relation on the translocating polyelectrolyte, we develop \BS{the first} correlation-dressed theory of ion and polymer transport \BS{accounting for the coexisting SC and WC electrostatics of the mono- and multivalent ion species, and the surface polarization forces originating from the dielectric contrast between the membrane matrix and the electrolyte}.

Within this EH theory, we identify a new polymer translocation mechanism triggered by electrostatic many-body interactions and membrane polarization forces. Namely, our formalism reveals that due to the attractive polarization forces governing the giant permittivity regime of \BS{surface}-coated membranes, added multivalent ions adsorbed by the neutral nanopore attract their monovalent counterions into the pore and reject their monovalent coions to the reservoir. Under the effect of the external voltage, this charge separation generates an EO counterion flow dragging the polymer at the velocity that can be tuned via the charge and amount of the added multivalent ions. The dielectrically generated translocation mechanism is therefore a central prediction for the transport of overall neutral or weakly charged molecules that cannot be controlled by an external voltage.

We characterize as well the effect of the electrolyte composition on the current signal generated by anionic polymer translocation events. We find that upon addition of \BS{multivalent cations such as spermidine ${\rm Spd}^{3+}$ or cobalt ${\rm Co}^{3+}$ charges} into the \BS{monovalent salt} reservoir, electrostatic many-body interactions replace the monovalent \BS{salt} counterions bound to the translocating polyelectrolyte by the \BS{multivalent} counterions of higher electric mobility. This charge discrimination amplifies the electric current signal by several factors. Considering the correlation between the ionic current strength and the single nucleotide resolution in biosequencing by translocation, the multivalent counterion-induced signal amplification is a key observation for nanopore-based sensing applications. 

\section{Theory}

We introduce here the electrostatic model of the polymer-electrolyte complex confined to a nanopore. Then, we combine this formalism with the hydrodynamic Stokes theory, and derive the polymer translocation velocity and charge current through the pore.

\subsection{Electrostatic model}

\begin{figure}
\includegraphics[width=1.\linewidth]{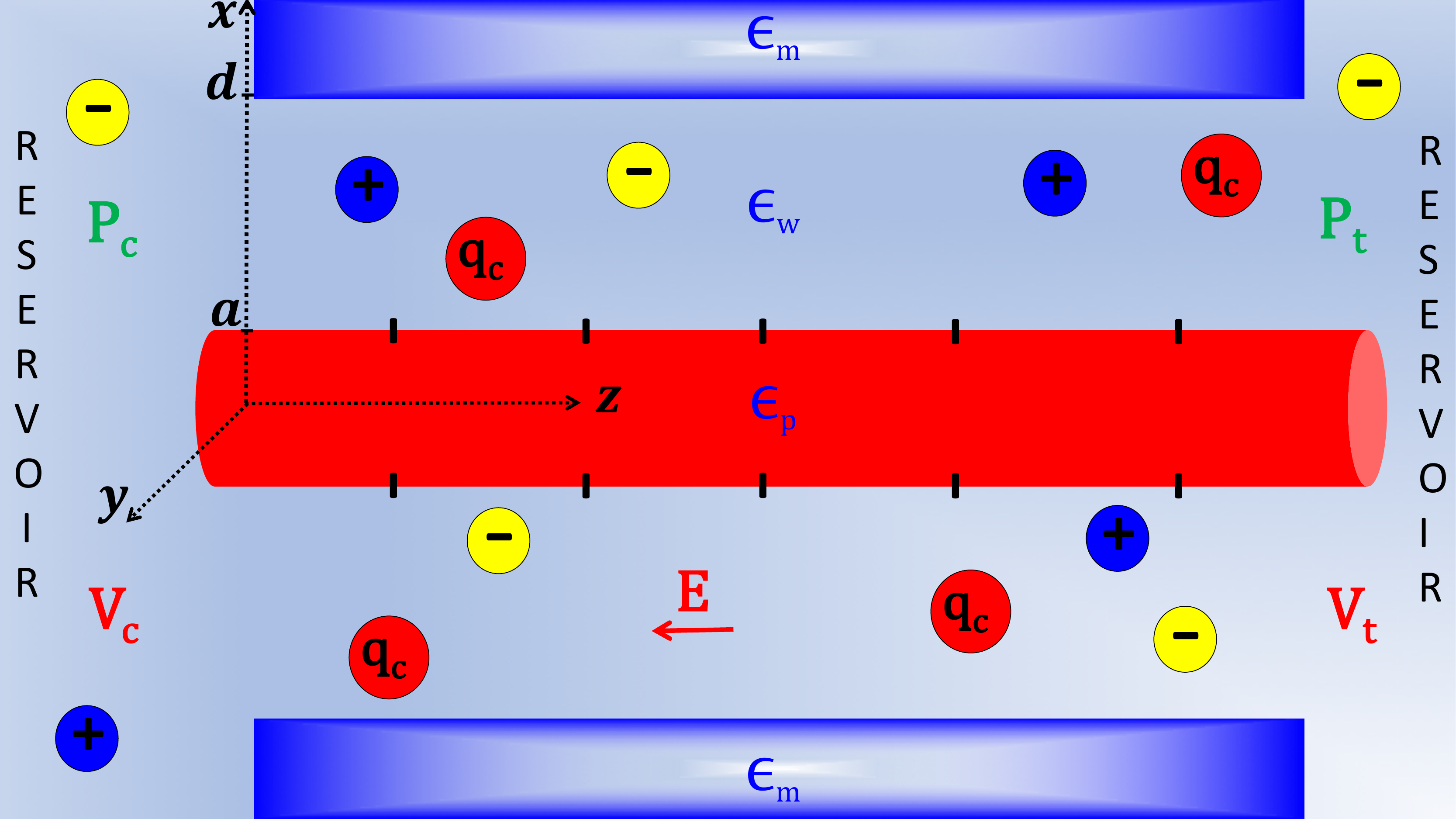}
\caption{(Color online) Cross sectional view of the cylindrical polymer of radius $a=1$ nm, negative surface charge density $-\sigma_{\rm p}$, and dielectric permittivity $\e_{\rm p}=2$, translocating through an interfacially neutral cylindrical pore of radius $d$ and permittivity $\e\m$~\cite{nt1}. \BS{The charged liquid of permittivity $\e_{\rm w}=80$ is composed of monovalent cations ${\rm C}^+$ and anions ${\rm A^-}$, and multivalent cations ${\rm M}^{q\ce}$. The pore experiences the applied electric field $\bE=-E\hu_z$ and pressure gradient $\Delta P=P\ce-P_{\rm t}$.}}
\label{fig1}
\end{figure}
Fig.~\ref{fig1} depicts the configuration of the polymer-electrolyte complex. The confined polymer is modeled as a rigid cylinder of radius $a=1$ nm, length $L_{\rm p}$, dielectric permittivity $\e_{\rm p}=2$, and surface charge density $\sigma_{\rm p}$. The field-driven translocation takes place along the axis of the uncharged cylindrical membrane nanopore of radius $d$ and length $L\m\ll L_{\rm p}$~\cite{nt1}. \BS{Hence, the off-axis and conformational polymer fluctuations are neglected. Moreover, as our model does not include the molecular details of CNTs, the alteration of the dielectric substrate permittivity by surface engineering techniques~\cite{Coating4,Coating5} will be approximately taken into account by the uniform membrane permittivity coefficient $\e_{\rm m}$.}

\BS{The charged liquid located between the translocating polymer and the nanopore is modeled within the dielectric continuum electrostatics. Thus, the medium is characterized by the piecewise dielectric permittivity distribution}
\be
\label{di1}
\e(r)=\e\m\theta(r-d)+\e_{\rm p}\theta(a-r)+\e_{\rm w}\theta(d-r)\theta(r-a),
\ee
where $\e_{\rm w}=80$ is the water permittivity. \BS{The confined liquid contains an ionic mixture composed of a monovalent salt solution such as KCl or NaCl}, with the ionic valencies $q_\pm=\pm1$ and bulk concentrations $n_{\rm b\pm}$, and a third multivalent ion species of valency $q\ce$ and concentration $n_{\rm bc}$. \BS{From now on, the monovalent cation, anion, and multivalent ion species will be denoted by the symbols ${\rm C}^+$, ${\rm A}^-$, and ${\rm M}^{q\ce}$, respectively.} 

\BS{In our formalism, the mobile ions are modeled as point charges interacting via the Coulomb potential. That is, ion specific effects associated with the extended charge structure and ionic hydration are neglected. Thus, the different ionic species in the liquid mixture are discerned in our model solely by their valency. The aforementioned approximations are based on the confluent results of nanofluidic~\cite{ExpHey} and polymer transport experiments~\cite{poltrex1,poltrex2,poltrex3} indicating the leading role of the ionic multivalency in the emergence of correlation effects on charge transport characterized in this work. Via the comparative discussion of these experiments and our theoretical results, this point will be further elaborated in the Conclusions.}

\BS{The technical details of the SCPB formalism used in our study can be found in Ref.~\cite{JPCB2020}. This} formalism treats the mono- and multivalent ions within the WC and SC electrostatics, respectively. In the Supporting Information, we review the \BS{key aspects of the} SCPB theory and adapt the formalism to the cylindrical geometry of the polymer-nanopore system. We show that the SCPB \SB{equation} for the electrostatic potential $\psi(r)$ has the form of a radial Poisson \SB{equation},
\be
\label{m1}
r^{-1}\partial_rr\partial_r\psi(r)+4\pi\ell_{\rm B}\left[\rho\ce(r)-\sigma_{\rm p}\delta(r-a_*)\right]=0,
\ee
with the Bjerrum length $\ell_{\rm B}=7$ {\AA}, the liquid charge density
\be
\label{m2}
\rho\ce(r)=\sum_{i=\pm,{\rm c}}q_in_i(r),
\ee
and the ion number density $n_i(r)$. In Eq.~(\ref{m1}), the radial coordinate of the polymer charges $a_*=a+l\s$ differs from the dielectric radii $a$ in Eq.~(\ref{di1}) by the characteristic length $l\s=2$ {\AA}. The  length $l\s$ takes into account the location of the surface charge groups within the aqueous medium as well as the hydrodynamic no-slip condition at the polymer and pore surfaces.

\subsection{Convective liquid and polymer velocities}

Under the externally applied voltage $\Delta V=V_{\rm t}-V_{\rm c}$ and the hydrostatic pressure gradient $\Delta P$, the convective velocity $u\ce(r)$ of the electrolyte experiencing the electric field $E=\Delta V/L\m$ obeys the Stokes \SB{equation} $\eta r^{-1}\partial_rr\partial_ru\ce-eE\rho\ce(r)+\Delta P/L\m=0$, with the water viscosity $\eta=8.91\times10^{-4}$ Pa s and the electron charge $e$. Combining the Stokes relation and the SCPB \SB{equation}~(\ref{m1}), and integrating, the convective liquid velocity follows as
\be\label{m4}
u\ce(r)=-\mu_{\rm e}E\psi(r)-\frac{\Delta Pr^2}{4\eta L\m}+c_1\ln r+c_2,
\ee
where we used the \SB{electrophoretic (EP)} mobility coefficient $\mu_{\rm e}=e/(4\pi\ell_{\rm B}\eta)$. In order to determine the integration constants $c_{1,2}$, we account for the force-balance relation $f_{\rm el}+f_{\rm sh}=0$ on the polymer, with the surface density of the electrostatic force $f_{\rm el}=e\sigma_{\rm p}E$ and the hydrodynamic shear force $f_{\rm sh}=\eta u\ce'(a_*)$. Imposing to Eq.~(\ref{m4}) the resulting constraint $u\ce'(a_*)=-\sigma_{\rm p}eE/\eta$, and the no-slip conditions $v_{\rm p}=u\ce(a_*)$ and $u\ce(d_*)=0$ where $d_*=d-l\s$, the liquid and polymer velocities follow in the form
\bea\label{m5II}
u\ce(r)&=&-\mu_{\rm e}E\left[\psi(r)-\psi(d_*)\right]\\
&&+\frac{\Delta P}{4\eta L\m}\left[d_*^2-r^2-2a_*^2\ln\left(\frac{d_*}{r}\right)\right],\nonumber\\
\label{m6II}
v_{\rm p}&=&-\mu_{\rm e}E\left[\psi(a_*)-\psi(d_*)\right]\\
&&+\frac{\Delta P}{4\eta L\m}\left[d_*^2-a_*^2-2a_*^2\ln\left(\frac{d_*}{a_*}\right)\right].\nonumber
\eea

\subsection{Electric and streaming currents}
\begin{figure}
\includegraphics[width=1\linewidth]{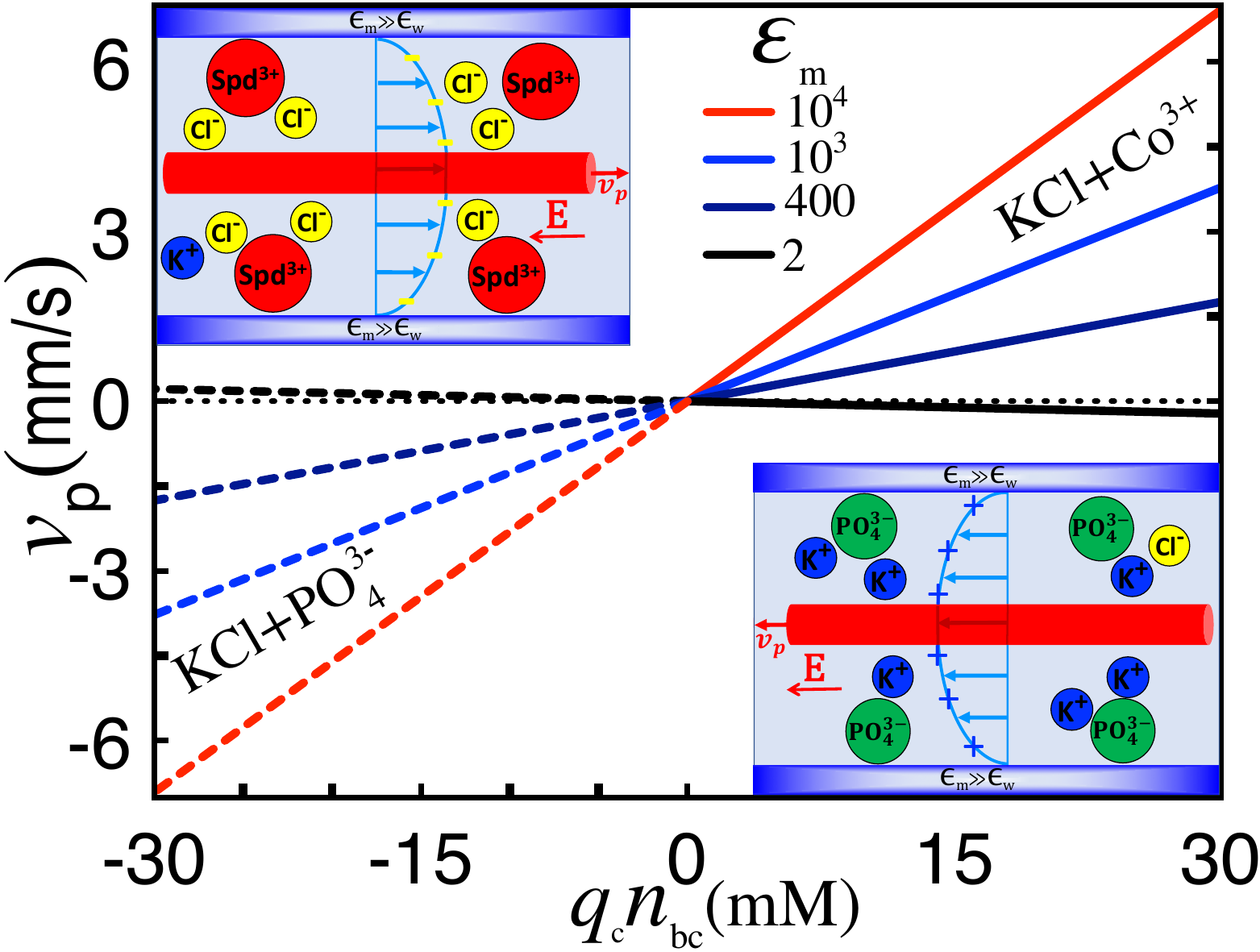}
\caption{(Color online) Translocation velocity~(\ref{m6II}) of a neutral protein ($\sigma_{\rm p}=0$) versus the multivalent charge concentration in an uncharged pore of radius $d=5$ nm and length $L\m=200$ nm. The voltage is $\Delta V=100$ mV. The electrolyte \BS{is composed of a monovalent salt mixed with trivalent cations ${\rm M^{3+}}$ (solid curves) or trivalent anions ${\rm M^{3-}}$ (dashed curves)}. The salt density is $n_{{\rm b}+}=0.1$ M. The insets illustrate the charge configuration in the surface-coated pore ($\e\m\gg\e_{\rm w}$) and the resulting EO flows dragging the neutral polymer.}
\label{fig2}
\end{figure}

The ionic current through the pore of cross-sectional area $S$ reads
\be\label{cr1}
I=e\int\mathrm{d}^2S\sum_{i=\pm,{\rm c}}q_in_i(r)u_i(r).
\ee
In Eq.~(\ref{cr1}), the total ion velocity $u_i(r)=u\ce(r)+u_{{\rm T}i}$ is given by the superposition of the convective hydrodynamic velocity~(\ref{m5II}), and the conductive velocity  $u_{{\rm T}i}=-\mu_i{\rm sign}(q_i)\Delta V/L\m$ including the ionic mobilities $\mu_i$~\cite{ionm}. Using the SCPB \SB{equation}~(\ref{m1}), and performing integrations by parts, the ionic current follows in the form of the linear response relation $I=G_{\rm V}\Delta V+G_{\rm P}\Delta P$, with the electric and streaming conductances respectively given by
\bea\label{gv}
G_{\rm V}&=&-\sum_{i=\pm,{\rm c}}|q_i|\mu_i\int_{a_*}^{d_*}\mathrm{d}rrn_i(r)\\
&&-\frac{2\pi\mu_ee}{L\m}\int_{a_*}^{d_*}\mathrm{d}rr\rho\ce(r)\left[\psi(r)-\psi(d_*)\right],\nonumber\\
\label{gp}
G_{\rm P}&=&\frac{e}{4\eta\ell_{\rm B}L\m}\left\{2\int_{a_*}^{d_*}\mathrm{d}rr\psi(r)-\left(d_*^2-a_*^2\right)\psi(d_*)\right\}\nonumber\\
&&+\frac{\pi ea_*\sigma_{\rm p}}{2\eta L\m}\left\{d_*^2-a_*^2-2a_*^2\ln\left(\frac{d_*}{a_*}\right)\right\}.
\eea
\begin{figure*}
\includegraphics[width=1\linewidth]{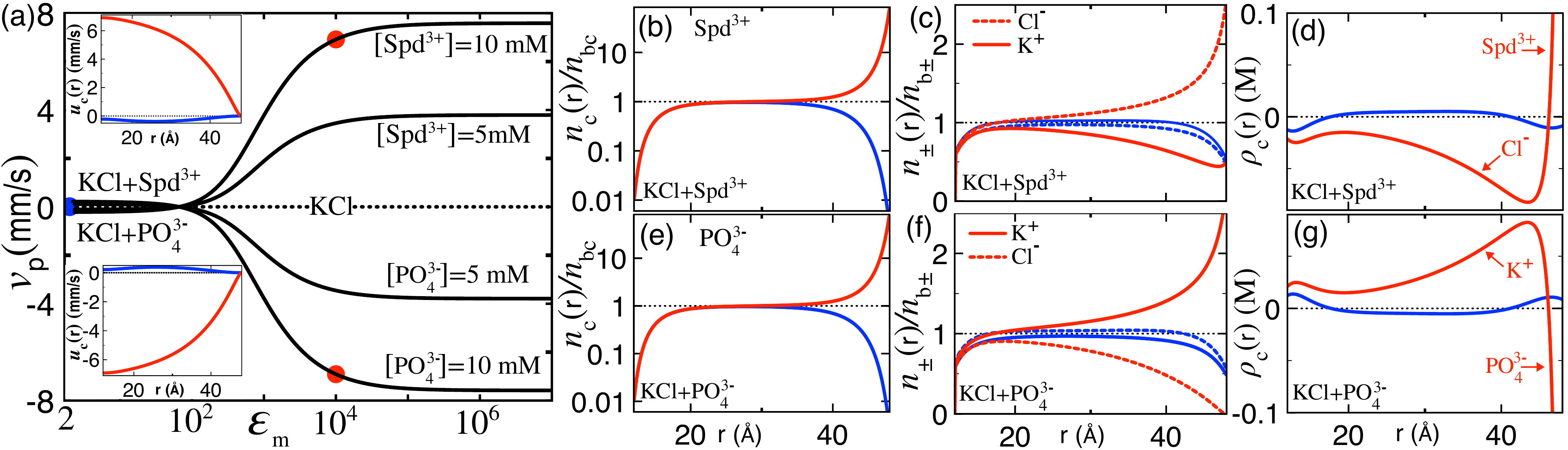}
\caption{(Color online) (a) Translocation velocity~(\ref{m6II}) versus the membrane permittivity (main plot) and the liquid velocity profile~(\ref{m5II}) (insets). (b)-(e) Densities of the multivalent ion species \BS{$\rm{M}^{3\pm}$}. (c)-(f) Densities of the monovalent \BS{cation species $\rm{C}^+$} (solid curves) and \BS{anion species $\rm{A}^-$} (dashed curves). (d)-(g) The net liquid charge density~(\ref{m2}). The electrolytes are \BS{salt+${\rm M^{3+}}$} (top plots) and \BS{salt+${\rm M^{3-}}$} (bottom plots). The membrane permittivity values are $\e\m =2$ (blue) and $\e\m =10^4$ (red). The remaining model parameters are the same as in Fig.~\ref{fig2}.}
\label{fig3}
\end{figure*}

The electric conductance~(\ref{gv}) is composed of the conductive flow component (first term) typically dominating the EO component (second term)~\cite{Gillespie2013}. Then, the first line of the streaming conductance~(\ref{gp}) is associated with the free ion contribution to the charge flow. In polymer-free open pores ($a_*=0$), the sign of the streaming current is set by the sign of the zeta potential term $-d_*^2\psi(d^*)$ bringing the major contribution~\cite{JPCB2020}. In the case of a translocating polymer, this free ion component is augmented by the second line of Eq.~(\ref{gp}) proportional to the pressure-driven polymer mobility in Eq.~(\ref{m6II}). This additional term brings in turn the ionic current contribution from the cations bound to the translocating molecule.

\section{Results and discussion}

\subsection{Effect of the membrane polarization on the polymer mobility}

We characterize here the effect of the membrane polarization forces on the electric field-driven polymer mobility in \BS{dielectric membrane} nanopores confining multivalent electrolytes. Thus, we turn off the pressure ($\Delta P=0$), and set the model parameters to the characteristic pore radius $d=5$ nm, length $L\m=200$ nm, and external voltage $\Delta V=100$ mV of the translocation experiments~\cite{exp2}.

\subsubsection{Generation of field-driven neutral polymer mobility}

Fig.~{\ref{fig2} displays the translocation velocity~(\ref{m6II}) of a neutral protein such as {\it avidin} at the point of zero charge ($p{\rm H}=9$)~\cite{Firnkes} against the multivalent charge density \BS{in the liquid mixtures including trivalent cations (${\rm M^{3+}}$) such as spermidine ${\rm Spd^{3+}}$ or cobalt ${\rm Co}^{3+}$ (solid curves), or trivalent anions (${\rm M^{3-}}$) such as phosphate ${\rm PO_4^{3-}}$ (dashed curves). At the dielectric  permittivity value $\e\m=2$ of} uncoated silicon membranes (black), the vanishingly small velocity exhibits an irrelevant dependence on the multivalent ion concentration. However, in the \BS{high permittivity regime $\e\m\gg\e_{\rm w}$ of engineered membrane pores}, added trivalent cations trigger a finite polymer mobility opposite to the external electric field ($v_{\rm p}>0$), and the addition of trivalent anions induces a mobility along the electric field ($v_{\rm p}<0$). Moreover, the mobility grows uniformly with the membrane permittivity ($\e\m\uparrow|v_{\rm p}|\uparrow$), and the polymer velocity evolves linearly with the multivalent charge density, i.e. $v_{\rm p}\propto q\ce n_{\rm bc}$.

The variation of the polymer velocity with the membrane permittivity is explicitly illustrated in the main plot of Fig.~\ref{fig3}(a). In pure symmetric \BS{monovalent salt such as KCl} (dotted curve), the velocity vanishes at all permittivity values. However, in the presence of multivalent ions (solid curves), the increment of the membrane permittivity from the insulator ($\e_{\rm m}<\e_{\rm w}$) to the conductor regime ($\e_{\rm m}>\e_{\rm w}$) drives the translocation velocity from a vanishingly small to a positive value for added trivalent cations, and to a negative value with added trivalent anions. This indicates that in the giant permittivity regime of \BS{surface}-coated pores, the valency and amount of the added multivalent ions can solely set the speed and direction of neutral or non-uniformly charged polymers that cannot be directly controlled by an external electric field.

\begin{figure}
\includegraphics[width=1\linewidth]{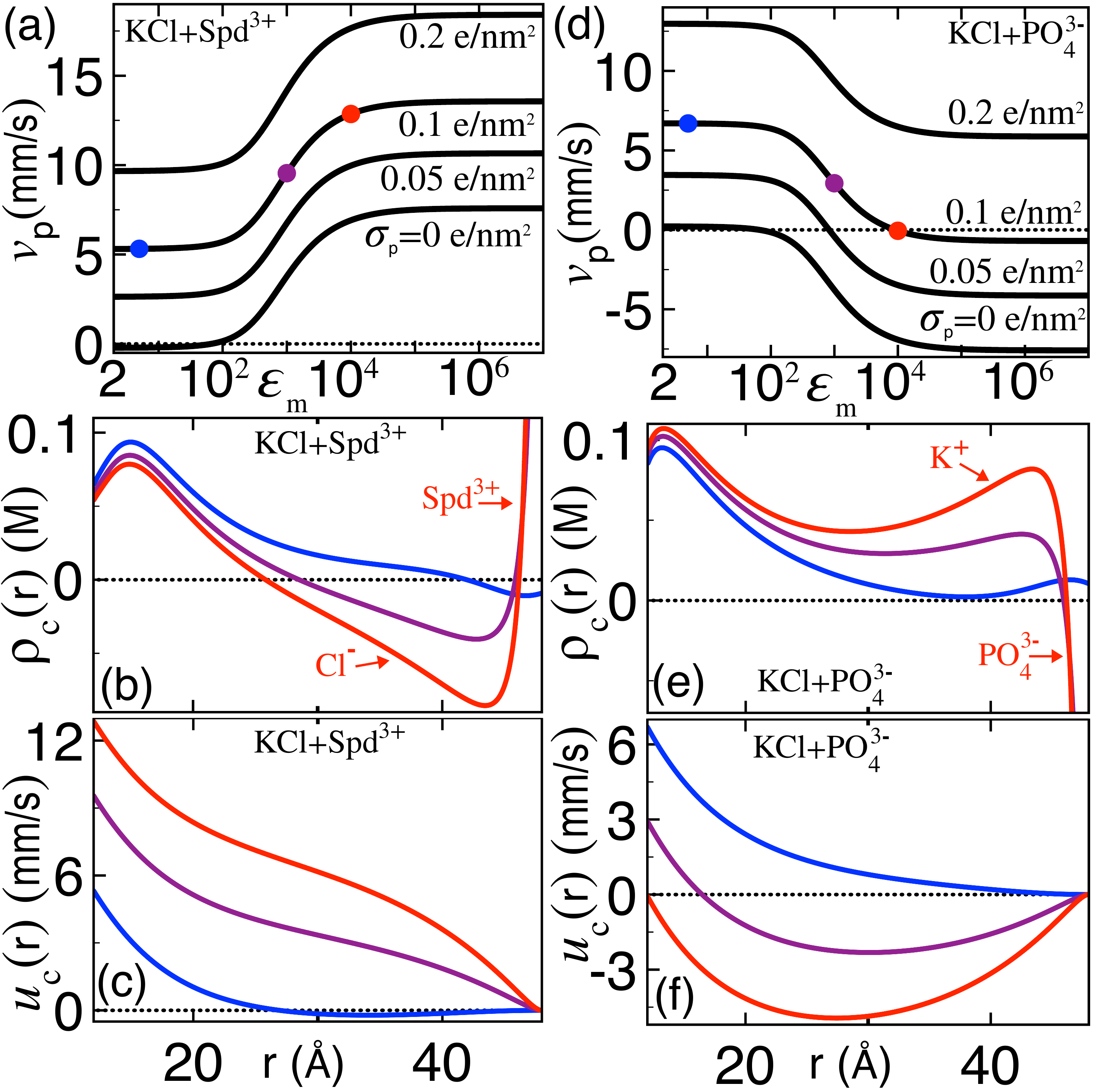}
\caption{(Color online) (a) Anionic polymer velocity~(\ref{m6II}), (b) liquid charge density~(\ref{m2}), and (c) liquid velocity~(\ref{m5II}) in the \BS{salt+${\rm M^{3+}}$} mixture. The polymer charge densities are given in the legend of (a). (d)-(f) The plots in (a)-(c) for the \BS{salt+${\rm M^{3-}}$} liquid. The ion densities are $n_{{\rm b}+}=0.1$ M and $n_{{\rm bc}}=0.01$ M. The permittivities of the colored curves are $\e\m =5$ (blue), $\e\m =10^3$ (purple), and $\e\m =10^4$ (red). The remaining model parameters are the same as in Fig.~\ref{fig2}.}
\label{fig4}
\end{figure}
Figs.~\ref{fig3}(b)-(d) illustrate for the \BS{liquid mixture including trivalent cations ($\rm M^{3+}$)} the mechanism triggering the neutral polymer mobility in terms of the charge densities. The membrane permittivity of each curve corresponds to the dot of the same color in Fig.~\ref{fig3}(a). Fig.~\ref{fig3}(b) shows that upon the rise of the permittivity from the insulator (blue) to the conductor regime (red), the inversion of the image-charge forces from repulsive to attractive leads to the adsorption of the \BS{trivalent cations} from the reservoir to the pore wall, i.e. $\e\m\uparrow n\ce(r)\uparrow$. In Fig.~\ref{fig3}(c), one sees that the resulting interfacial \BS{${\rm M^{3+}}$} layer acting as a cationic surface charge attracts \BS{monovalent anions (${\rm A}^-$)} into the pore and excludes \BS{the monovalent cations (${\rm C}^+$)} from the pore \BS{to the reservoir, i.e.} $\e\m\uparrow n_-(r)\uparrow n_+(r)\downarrow$~\cite{nt3}. According to Fig.~\ref{fig3}(d), in the central pore region occupied by the polymer, this charge separation gives rise to a \BS{monovalent anion}-rich liquid of \BS{net} negative charge ($\rho\ce(r)<0$). The coupling of this anionic liquid to the external field sets an EO flow dragging the polymer opposite to the field, i.e. $u\ce(r)>0$ and $v_{\rm p}>0$ (see the top insets of Figs.~{\ref{fig2} and~\ref{fig3}(a)).

Figs.~\ref{fig3}(e)-(g) show that for the \BS{electrolyte including trivalent anions ($\rm M^{3-}$)}, the transition from the insulator to the giant permittivity regime reverses this transport picture. Namely, the trivalent anions driven by the attractive polarization forces bring extra \BS{monovalent salt cations} into the pore and reject the \BS{monovalent salt anions} from the pore \BS{to the reservoir}. This ion discrimination leads to a \BS{monovalent cation}-rich liquid ($\rho\ce(r)>0$) and a cationic EO flow dragging the neutral polymer along the electric field, i.e. $u\ce(r)<0$ and $v_{\rm p}<0$ (see the bottom insets of Figs.~{\ref{fig2} and~\ref{fig3}(a)).

\subsubsection{Inverting the mobility of weakly anionic polymers} 
\label{weak}

Having characterized the emergence of neutral polymer mobility from the polarization-induced EO flows, we scrutinize now the collective effect of these flows and the EP drag on the translocation of weakly anionic polymers such as deprotonated avidin molecules~\cite{Firnkes}. Fig.~\ref{fig4}(a) shows that in the liquid \BS{containing trivalent cations}, the transition from the insulator to the giant permittivity regime amplifies the positive EP polymer velocity. The mechanism driving this effect is displayed in the charge density and liquid velocity profiles of Figs.~\ref{fig4}(b)-(c). In the insulating regime of \BS{silicon-based} membranes (blue), the cation cloud ($\rho\ce(r)>0$) dragged by the electrophoretically translocating anionic polymer moves opposite to the electric field ($u\ce(r)>0$). Then, in \BS{the large permittivity regime of engineered membrane nanopores} governed by attractive polarization forces (purple and red), the \BS{monovalent anions} attracted by the interfacially adsorbed \BS{trivalent cations} invert the liquid charge from positive ($\rho\ce(r)>0$) to negative ($\rho\ce(r)<0$). The resulting anionic EO flow moving opposite to the external electric field enhances the positive liquid and polymer velocities, i.e. $\e\m\uparrow u\ce(r)\uparrow v_{\rm p}\uparrow$.

Fig.~\ref{fig4}(d) indicates that in the \BS{mixture containing trivalent anions}, the rise of the membrane permittivity from the insulator to the conductor regime can suppress or even invert the EP polymer velocity. To shed light on this peculiarity, we note that according to Fig.~\ref{fig4}(e), the \BS{monovalent} cations attracted by the interfacial \BS{trivalent anion} layer enhance the positive density of the counterion liquid around the anionic polymer. The coupling of this strongly cationic liquid to the electric field sets an EO current parallel with the field.  Fig.~\ref{fig4}(f) shows that the drag of this EO flow opposing the EP force inverts the sign of the liquid and polymer velocities from positive to negative, i.e. $\e\m\uparrow u\ce(r)\downarrow v_{\rm p}\downarrow$. In particular, at the membrane permittivity $\e\m=10^4$ (red), the exact compensation of the EP drift by the EO flow stops entirely the translocation, i.e. $u\ce(a_*)=v_{\rm p}=0$.

\begin{figure}
\includegraphics[width=1\linewidth]{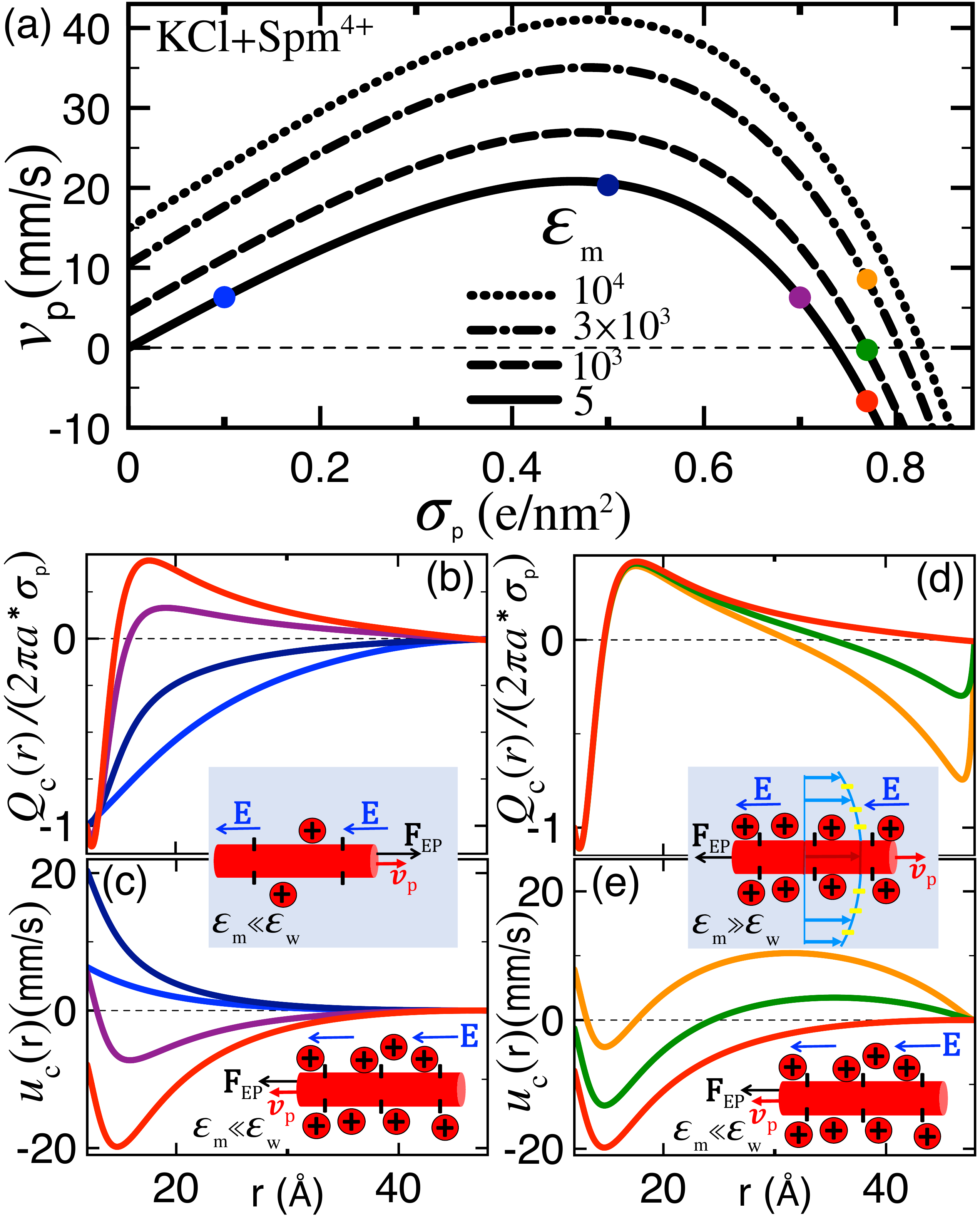}
\caption{(Color online) (a) Translocation velocity~(\ref{m6II}) in the \BS{salt+${\rm M}^{4+}$} mixture versus the polymer charge strength at the bulk \BS{${\rm M}^{4+}$}  density $n_{\rm bc}=1$ mM and the membrane permittivities in the legend. (b)-(d) Cumulative charge~(\ref{cumch}) and (c)-(e) convective liquid velocity~(\ref{m5II}) at the polymer charge densities of the circles with the same color in (a): $\sigma_{\rm p}=0.1$ (blue), $0.5$ (navy), $0.7$ (purple), and $0.77$ ${\rm e}/{\rm nm}^2$ (red, green, and orange). The other parameters are the same as in Fig.~\ref{fig2}.}
\label{fig5}
\end{figure}
\subsubsection{Mobility inversion of strongly anionic biopolymers}
\label{invsc}

In the \BS{experimentally studied} case of strongly charged biopolymers such as DNA translocating in the highly multivalent KCl+${\rm Spm}^{4+}$ mixture~\cite{poltrex1,poltrex2,poltrex3}, the polarization-driven EO flows are expected to act together with the strong-coupling correlations governing the vicinity of the polymer. With the aim to investigate the outcome of this coupling, we reported in Fig.~\ref{fig5}(a) the surface charge dependence of the polymer velocity at various membrane permittivities \BS{for the electrolyte mixture composed of monovalent salt and tetravalent cations (${\rm M}^{4+}$)}. For moderate polymer charges $\sigma_{\rm p}\lesssim0.5$ ${\rm e}/{\rm nm}^2$, the increase of the charge density amplifies the electrostatic force $F_{\rm el}=e\sigma_{\rm p}E$ and enhances steadily the polymer velocity, i.e. $\sigma_{\rm p}\uparrow v_{\rm p}\uparrow$. However, in the strong polymer charge regime $\sigma_{\rm p}\gtrsim0.5$ ${\rm e}/{\rm nm}^2$,  this linear response behavior breaks down; therein, the rise of the polymer charge reduces the translocation velocity $\sigma_{\rm p}\uparrow v_{\rm p}\downarrow$ and inverts its sign from positive to negative. 

The \BS{inverted} velocity regime observed in experiments~\cite{poltrex1,poltrex2,poltrex3} and simulations~\BS{\cite{aks2} involving various multivalent charge species} corresponds to an unconventional transport picture characterized by an anionic polymer translating parallel with the electric field. Figs.~(\ref{fig5})(b)-(c) illustrate the underlying EH mechanism in terms of the cumulative charge~\cite{nt2}
\be\label{cumch}
Q\ce(r)\equiv2\pi\int_{a_*}^r\mathrm{d}r'r'\left[\rho\ce(r')-\sigma_{\rm p}\delta(r'-a_*)\right]=-\frac{r\psi'(r)}{2\ell_{\rm B}}
\ee
and the liquid velocity~(\ref{m5II}) \BS{in the low permittivity regime of uncoated neutral pores ($\e_{\rm m}=2$)} where EO flows are absent. In the case of moderately charged polymers (blue and navy), the partial screening of the polymer charges by the surrounding counterions results in a uniformly negative total charge ($Q\ce(r)<0$). This generates a liquid flow and a polymer mobility opposite to the field $\bE$ ($u\ce(r)>0$ and $v_{\rm p}>0$). Upon the increment of the polymer charge \BS{density} to $\sigma_{\rm p}=0.77$ ${\rm e}/{\rm nm}^2$ (red), enhanced correlations between the \BS{multivalent cations} and polymer charges lead to the overcompensation of the latter by \BS{these} counterions, and to the inversion of the cumulative charge from negative to positive ($Q\ce(r)>0$). The coupling of the field $\bE$ to this cationic polymer-liquid complex reverses the EP force, reorienting the liquid and polymer velocities along the field ($u\ce(r)<0$ and $v_{\rm p}<0$). This mechanism is depicted in the insets of Figs.~\ref{fig5}(b)-(c).

In Figs.~(\ref{fig5})(a)-(b), one notes that the intermediate charge strength $\sigma_{\rm p}=0.7$ ${\rm e}/{\rm nm}^2$ (purple) gives rise to CI ($Q\ce(r)>0$) without the polymer mobility reversal ($v_{\rm p}>0$). To elucidate this point, we combine Eqs.~(\ref{m5II}) and~(\ref{cumch}) to obtain the macroscopic force-balance relation
\be\label{fb}
eQ\ce(r)LE=2\pi rL\eta u'(r)
\ee
equating the EP force (l.h.s.) and the hydrodynamic shear force (r.h.s.) acting on the cylindrical liquid volume $V(r)=2\pi rL$. Eq.~(\ref{fb}) and Figs.~\ref{fig5}(b)-(c) show that CI at $r=r^*$ results in a liquid velocity minimum ($u'\ce(r^*)=0$) accompanied with an inverted EO flow ($u\ce(r)<0$). However, the close inspection of Fig.~\ref{fig5}(c) also shows that the flow inversion causes the reversal of the polymer velocity $v_{\rm p}=u\ce(a_*)$ exclusively if the positive slope $u'(r)$ is large enough at $r>r^*$. Thus, according to Eq.~(\ref{fb}), the occurrence of polymer mobility inversion requires a strong enough reversed positive charge at $r>r^*$.

\begin{figure}
\includegraphics[width=1.\linewidth]{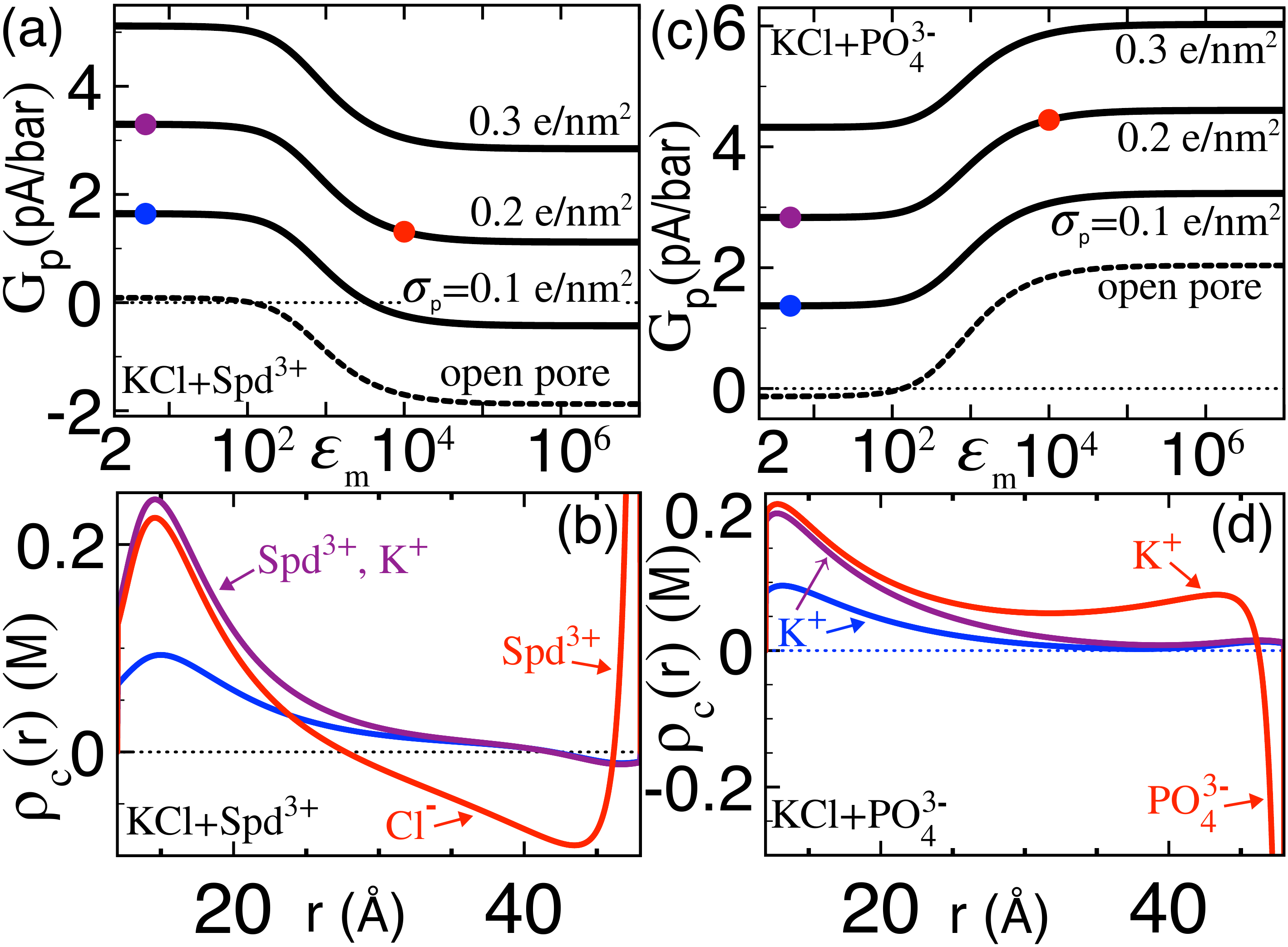}
\caption{(Color online) (a) Streaming conductance~(\ref{gp}) and (b) liquid charge density~(\ref{m2}) in the \BS{salt+${\rm M^{3+}}$} mixture. The polymer charge densities are given in the legend of (a). (c)-(d) The plots in (a)-(b) for the \BS{salt+${\rm M^{3-}}$} liquid. The ion densities are $n_{{\rm b}+}=0.1$ M and $n_{{\rm bc}}=0.01$ M. The curves in (b)-(d) are plotted at the parameters of the dots with the same color in (a)-(b). The pore length is $L\m=34$ nm~\cite{e6}. The remaining model parameters are the same as in Fig.~\ref{fig2}.}
\label{fig6}
\end{figure}

Finally,  Fig.~\ref{fig5}(a) shows that in the inverted mobility phase, the increment of the membrane permittivity from the insulator (red) to the giant permittivity regime (green and orange) reverses the polymer velocity from negative back to positive. The mechanism behind the cancellation of mobility inversion is illustrated in Figs.~\ref{fig5}(d)-(e). One notes that the attractive polarization forces emerging at high permittivities do not affect the cumulative charge $Q\ce(r)$ and the associated EP force on the polymer-counterion complex at $r\lesssim 20$ {\AA}. However, in the outer liquid layer at $r\gtrsim 20$ {\AA}, the  \BS{monovalent anions} attracted by the interfacially adsorbed \BS{tetravalent cations} turn the cumulative charge from weakly positive to strongly negative. The resulting anionic flow of positive velocity drags the charge inverted polymer oppositely to the electric field ($v_{\rm p}>0$). Thus, \BS{in surface-coated pores located in the giant permittivity regime}, the \BS{tetravalent cation}-induced EO flows are expected to suppress the EP mobility inversion triggered by the same \BS{tetravalent} charges in uncoated pores~\cite{poltrex1,poltrex2}. The complex EH picture behind this prediction is illustrated in the insets of Figs.~\ref{fig5}(d)-(e).

\begin{figure*}
\includegraphics[width=1.\linewidth]{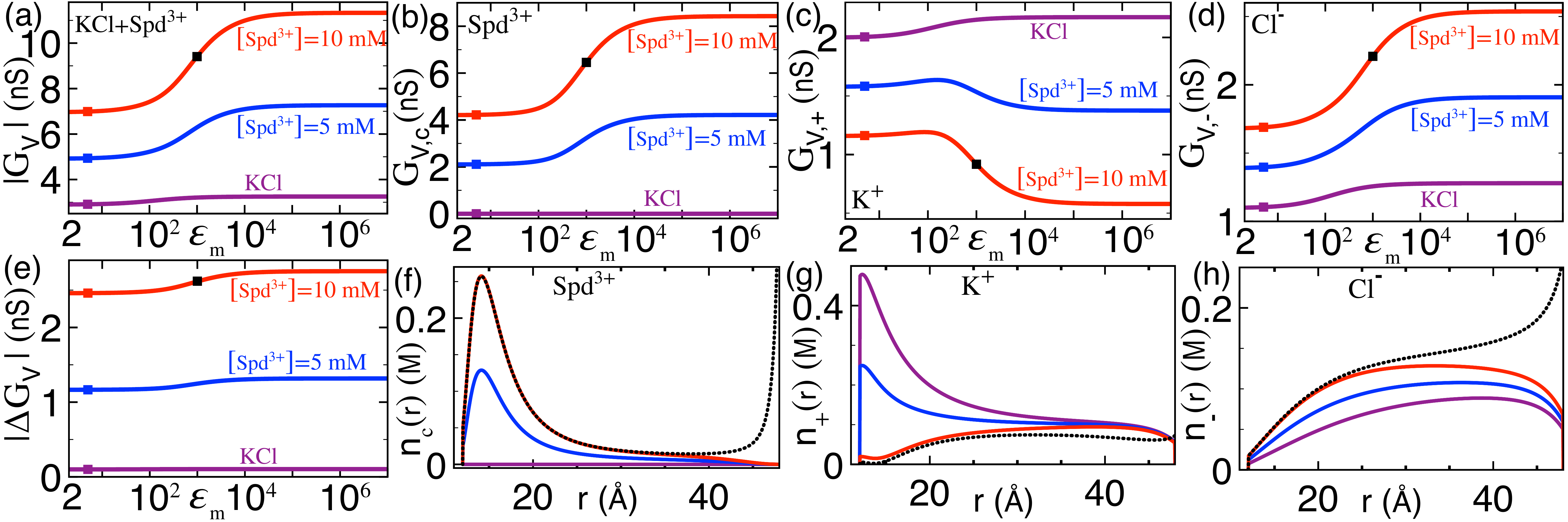}
\caption{(Color online) (a) Electric conductance~(\ref{gv}), (b)-(d) individual ion conductivities in Eq.~(\ref{gvcom}), and (e) current signal $\Delta G_{\rm V}=G_{\rm V}(a)-G_{\rm V}(a\to0)$ generated by the anionic polymer of charge density $\sigma_{\rm p}=0.35$ ${\rm e}/{\rm nm}^2$ translocating in the \BS{salt+${\rm M^{3+}}$} solution. (f) ${\rm M}^{3+}$, (g) ${\rm C}^{+}$,  and (h) ${\rm A}^{-}$ density at the parameters corresponding to the square symbols of the same color in (a)-(e). The remaining parameters are the same as in Fig.~\ref{fig6}.}
\label{fig7}
\end{figure*}

\subsection{Polarization effects on the current signal}

The dependence of the sequencing resolution on the strength of the current signal requires the identification of the experimentally tunable system features maximizing the magnitude of the current. Motivated by this point, we scrutinize here the influence of membrane coating and ion multivalency on the current signal generated by pressure and voltage-driven translocation events.

\subsubsection{Streaming current signal}

Fig.~\ref{fig6}(a) displays the streaming current~(\ref{gp}) induced by pressure-driven translocation versus the membrane permittivity in the \BS{liquid containing trivalent cations}. This figure should be interpreted together with the charge density plot in Fig.~\ref{fig6}(b). The latter indicates that during translocation through insulating membranes (blue and purple),  the counterions attracted by the anionic polymer form a cationic layer around the molecule ($\sigma_{\rm p}\uparrow\rho\ce(r)\uparrow$). In Fig.~\ref{fig6}(a), the comparison of the dashed (open pores) and solid curves (blocked pores) shows that this cation excess generates a positive current signal ($\sigma_{\rm p}\uparrow G_{\rm p}\uparrow$) corresponding to the second term of Eq.~(\ref{gp}). 

As the permittivity rises from the insulator (purple) to the giant permittivity regime (red), the cation-driven current is substantially suppressed ($\e\m\uparrow G_{\rm p}\downarrow$). This effect is driven by the charge separation mechanism scrutinized in Section~\ref{weak}. Namely, Fig.~\ref{fig6}(b) shows that in the conductor regime $\e_{\rm m}\gg\e_{\rm w}$, the \BS{trivalent cations} adsorbed by the polarization forces bring the \BS{monovalent anions} close to the pore wall. The resulting anionic liquid layer suppresses the positive charge of the streaming flow ($\e\m\uparrow\rho\ce(r)\downarrow$) and weakens the cationic current through the nanopore. 

Fig.~\ref{fig6}(a) also shows that in the giant permittivity regime $\e_{\rm m}\gg\e_{\rm w}$, the same \BS{monovalent anions} generate a net negative streaming current through polymer-free open pores. The seemingly counterintuitive \SB{presence} of a streaming \SB{current} in an uncharged pore has been characterized in our recent work on ion transport thorough nanoslits~\cite{JPCB2020}. The occurrence of these \SB{currents} in the present nanopore geometry implies that in \BS{mixed electrolytes including trivalent cations}, anionic polymer penetration into a \BS{surface-coated pore} can reverse the streaming current from a negative to a positive value.

Finally, Figs.~\ref{fig6}(c)-(d) illustrate the streaming current and the charge partition in the \BS{electrolyte containing trivalent anions}. In the regime of insulating membranes (blue and purple), the attraction of the \BS{monovalent} counterions, and the repulsion of the \BS{mono- and trivalent} coions by the polymer charges generates a positive current signal. Moving to the giant permittivity regime (red), the extra \BS{monovalent cations} attracted by the interfacially adsorbed \BS{trivalent anions} positively add to the counterions brought by the translocating polyelectrolyte. This strengthens the positive mobile charge density and the streaming current through the pore, i.e. $\e\m\uparrow\rho\ce(r)\uparrow G_{\rm p}\uparrow$. 

\subsubsection{Amplification of the electric current signal by \BS{multivalent counterions}}

We predict herein the amplification of the electric current signal by multivalent counterions in anionic polymer translocation. Figs.~\ref{fig7}(a)-(e) illustrate the electric conductance~(\ref{gv}), the individual ion conductivities
\be\label{gvcom}
G_{{\rm V},i}=\frac{2\pi e}{L\m}|q_i|\mu_i\int_{a_*}^{d_*}\mathrm{d}rrn_i(r)
\ee
with $i=\{\pm,{\rm c}\}$, and the conductivity change $\Delta G_{\rm V}=G_{\rm V}(a)-G_{\rm V}(a\to0)$ upon polymer penetration versus the membrane permittivity $\e_{\rm m}$ and \BS{trivalent counterion} concentration. One notes that in insulating membrane pores ($\e_{\rm m}\ll\e_{\rm w}$), added \BS{trivalent} cations enhance the current by several factors ($n_{\rm bc}\uparrow |G_{\rm V}|\uparrow$) and amplify the net current signal from an insignificant value to a finite magnitude roughly proportional to the \BS{trivalent counterion} concentration, \BS{i.e.} $|\Delta G_{\rm V}|\propto n_{\rm bc}$.

Figs.~{\ref{fig7}(f)-(h) indicate that the signal amplification is set by the electrostatic many-body interactions replacing the \BS{monovalent} counterions around the polymer by the \BS{trivalent} counterions of higher electric mobility. Indeed, one sees that the increment of the bulk \BS{trivalent cation} concentration from the purple to the blue and red curves enhances the density of the \BS{trivalent counterion} layer around the \BS{polymer}. However, these cations binding to the polymer charges exclude the \BS{monovalent counterions} from the polymer surface and attract the \BS{monovalent coions} into the pore, i.e. $n_{\rm bc}\uparrow n\ce(r)\uparrow n_+(r)\downarrow n_-(r)\uparrow$. Figs.~{\ref{fig7}(b)-(d) show that this charge separation enhances the \BS{${\rm M}^{3+}$} and \BS{${\rm A}^{-}$} conductivities and lowers the \BS{${\rm C}^{+}$} conductivity of the nanopore, i.e. $n_{\rm bc}\uparrow G_{{\rm V},{\rm c}}\uparrow G_{{\rm V},-}\uparrow G_{{\rm V},+}\downarrow$. Thus, in contrast with the \BS{purely monovalent salt solutions} where the current signal is governed by the excess of \BS{monovalent} counterions~\cite{e6,the16}, the signal strength in the \BS{salt+${\rm M^{3+}}$} mixture is set by the excess of \BS{trivalent} counterions accompanied with the \BS{monovalent counterion} exclusion from the pore. This elektrokinetic charge discrimination is precisely responsible for the signal amplification in Fig~\ref{fig7}(e).

Fig.~{\ref{fig7}(a) shows that in the \BS{salt+${\rm M}^{3+}$} liquid, the rise of the membrane permittivity from the insulator to the giant permittivity regime leads to a further amplification of the electric current ($\e\m\uparrow |G_{\rm V}|\uparrow$). The underlying charge configuration is displayed in Figs.~{\ref{fig7}(f)-(h). One sees that upon the transition from the insulator (red) to the conductor regime (black), the emerging polarization forces give rise to an additional \BS{${\rm M}^{3+}$} adsorption peak close to the pore wall. This interfacial \BS{trivalent cation} layer enhances the \BS{monovalent anion} density and leads to further \BS{monovalent cation}  depletion from the pore, i.e. $\e\m\uparrow n\ce(r)\uparrow n_-(r)\uparrow n_+(r)\downarrow$. Figs.~\ref{fig7}(b)-(d) show that the corresponding charge discrimination results in the enhancement of the \BS{${\rm M}^{3+}$} and \BS{${\rm A}^{-}$} conductivities and the reduction of the \BS{${\rm C}^{+}$} conductivity of the translocated pore, i.e. $\e\m\uparrow G_{{\rm V},{\rm c}}\uparrow G_{{\rm V},-}\uparrow G_{{\rm V},+}\downarrow$. Thus, \BS{the giant permittivity condition reached in surface-coated membranes strengthens} the current amplification induced by multivalent counterions.

\section{Conclusions}

The accurate depiction of polymer translocation in mixed solutions requires the consideration of the multiple electrostatic coupling strengths associated with the distinct ionic valencies in the liquid. Within an EH theory combining the hydrodynamic Stokes relation and the electrostatic SCPB formalism fulfilling this requirement, we characterized the coupled effect of the ion multivalency and the membrane polarization forces on driven polymer translocation through \BS{dielectric membrane nanopores located in the giant permittivity regime}.

Our formalism \BS{predicts} that multivalent charge addition into a \BS{monovalent salt} solution can solely trigger the voltage-driven mobility of {\it neutral polymers} \BS{through engineered membrane pores of giant permittivity}. The effect is induced by the attractive membrane polarization forces adsorbing the multivalent ions into the nanopore. The resulting multivalent charge excess sets a monovalent ion separation and generates an  EO counterion flow dragging the polymer through the pore. Due to the linear dependence of the polymer velocity on the charge and amount of the added multivalent ions, this EH process can be efficiently used to tune the mobility of overall neutral proteins that cannot be directly controlled by an electric field. This indicates that the polarization-driven polymer transport mechanism is a prediction of high relevance for nanopore-based biosensing technologies.

In the case of {\it anionic polymer} translocation, we showed that the multivalent cation-induced EO flows enhance the EP polymer mobility whereas multivalent anion addition reduces and even inverts the polymer velocity. Applying the transport formalism to the \BS{charge configuration of the translocation experiments conducted with strongly charged biopolymers in the KCl+${\rm Spm^{4+}}$ mixture}, we found that the \BS{tetravalent cation}-induced EO flows in surface-coated pores suppress the experimentally observed DNA mobility reversal triggered by the same \BS{tetravalent} charges in uncoated pores~\cite{poltrex1,poltrex2}. 

We carried out as well the first characterization of the surface polarization effects and ion multivalency on the current signal generated by translocation events. In pressure-driven translocation through \BS{surface}-coated pores \BS{with giant membrane permittivity}, due to the short range of the attractive polarization forces and the reduced surface velocity of the Poiseuille flow, the multivalent ions adsorbed to the pore wall affect the streaming current signal as overall inert interfacial charges. Consequently, their net contribution to the streaming current is opposite to their charge; \BS{trivalent anion} addition into \BS{monovalent salt} strengthens the positive current while \BS{trivalent cation} addition weakens or reverses the current signal from a positive to a negative value.  

In contrast with pressure-driven translocation, \BS{trivalent cation} addition into \BS{monovalent salt} amplifies the electric current signal generated by voltage-driven translocation events by several factors. The amplification of the current is set by the electrostatic many-body interactions replacing the monovalent polymer counterions by the multivalent cations of higher mobility. Moving to the giant permittivity regime of \BS{engineered membrane} pores, the attractive polarization forces lead to further \BS{trivalent cation} adsorption and result in an additional enhancement of the electric current.  Considering the importance of the signal magnitude for the precision of polymer sequencing by translocation, the current amplification by ion multivalency and \BS{giant membrane permittivity} is a central prediction for biosequencing purposes. 

\BS{In this work, we developed the first polymer transport formalism able to account for the coexisting WC and SC correlations in mixed electrolytes, and the polarization forces present in dielectric membrane pores. Due to the substantial complexity of the polymer translocation process, our theory naturally includes certain approximations. First, our stiff polyelectrolyte model neglects the conformational polymer fluctuations. For double-stranded DNA molecules in mono- and divalent bulk liquids, this approximation is valid for molecular lengths up to the persistence length $\ell_{\rm p}\sim30-55$ nm~\cite{pers}. However, it should be noted that the length $\ell_{\rm p}$ can be reduced by the presence of higher valency counterions, whereas the nanopore confinement limiting the polymer fluctuations is expected to extend the validity regime of the stiff polymer assumption. Therefore, the characterization of the rigid polymer approximation in the confined pore geometry will require the evaluation of these opposing effects by extensive numerical simulations.}

\BS{In addition, our electrolyte model is based on the dielectric continuum and the point-charge approximations neglecting the extended structure of the water molecules and the ion species, respectively, as well as the inhomogeneous hydration of ions by the solvent molecules. Furthermore, as our membrane model does not account for the molecular details of the engineered membranes, the alteration of the membrane permittivity was taken into account by a uniform membrane permittivity coefficient. The net effect of the corresponding approximations can be evaluated in future works via the explicit inclusion of the ionic and solvent charge structure~\cite{nonloc}, the dipolar membrane structure~\cite{DipFix}, and surface adsorption effects specific to the corresponding membrane material.} 

\BS{We emphasize that despite the aforementioned approximations, our transport theory has been so far able to reproduce and characterize two different experimentally observed correlation-driven transport phenomena, indicating that the formalism captures the essential features of the correlation effects under consideration. Namely, in Section~\ref{invsc}, we showed that the SC correlations mediated by the multivalent charges can solely trigger the EP polymer mobility inversion previously observed in polymer transport experiments~\cite{poltrex1,poltrex2,poltrex3}. Then, in our earlier work on the application of the electrostatic SCPB formalism to ion transport through nanoslits~\cite{JPCB2020}, we found that the occupation of the interfacial no-slip layer by multivalent charges is the mechanism behind the formation of like-charge streaming currents identified in early nanofluidic experiments by Heyden et al.~\cite{ExpHey}.}

\BS{The confluence between our theoretical results based on the point-charge model, and the observations made in these different experimental setups and charge configurations can be explained by noting the common feature of these experiments. Indeed, the EP polymer transport experiments of Refs.~\cite{poltrex1,poltrex2,poltrex3} as well as numerical simulations by Luan and Aksimentiev~\cite{aks3} show that the polymer mobility inversion occurs similarly upon the addition of tri- and higher valency charges characterized by radically different molecular structures, such as the trivalent cobalt hexamine and spermidine cations, the quadrivalent spermine molecules, the multivalent polylysine molecules of tri, four, and eight cationic charges, and the chitosan molecules. Moreover, in the nanofludic experiments of Heyden et al. conducted with negatively charged slits~\cite{ExpHey}, the streaming current inversion from positive to negative was shown to be triggered equally by the addition of divalent magnesium and calcium ions, as well as with trivalent cobalt sepulchrate molecules. The uniquely shared feature of these substantially different charge species inducing the same correlation-driven transport mechanism is the ionic multivalency. This indicates that the charge multivalency at the basis of our formalism plays the leading role in the emergence of correlation effects in nanoconfined charge transport.}

The dielectrically generated EO flows predicted \BS{in our work} are \BS{equally} expected to facilitate polymer capture from the reservoir into the pore. Due to the relevance of the polymer capture speed for serial DNA sequencing, future works should investigate the effect of these flows on the diffusion-limited polymer approach prior to translocation~\cite{mut1}. We also emphasize that our conclusions can be corroborated by current polymer translocation techniques involving surface-coated membrane nanopores~\cite{Coating1,Coating2,Coating3} and multivalent electrolytes~\cite{poltrex1,poltrex2}. \BS{The direct confrontation of our theory with these experiments will enable to identify the limitations of our model approximations and to determine the required extensions.}

\smallskip
\appendix
\section{SCPB formalism}
\label{for}

We review here the SCPB formalism developed in Ref.~\cite{JPCB2020} and adapt the theory to the cylindrical geometry of a polyelectrolyte with radius $a$ translocation through a nanopore of radius $d$. Within the SCPB theory, the monovalent background salt of the electrolyte mixture is treated at the weak electrostatic coupling level. Then, the high valency component characterized by SC interactions is taken into account within a virial approximation known to be equivalent to a SC treatment~\cite{NetzSC}.

The net average potential in the system is given by
\be\label{sup}
\psi(\br)=\phi\s(\br)+\phi\ce(\br),
\ee
with the salt-dressed potential $\phi\s(\br)$ and the multivalent ion-induced potential $\phi\ce(\br)$ solving the coupled equations
\bea\label{SCPB1}
&&\hspace{-3mm}\frac{k_{\rm B} T}{e^2}\nabla\cdot\e(\br)\nabla\phi\s(\br)+\sum_{i=\pm}q_in_i(\br)+\sigma(\br)=-\frac{\kappa^2(\br)}{4\pi\ell_{\rm B}}\phi\ce(\br),\nonumber\\
&&\\
\label{mp0}
&&\frac{k_{\rm B} T}{e^2}\left[\nabla\e(\br)\nabla-\e(\br)\kappa^2(r)\right]\phi\ce(\br)=-q\ce n\ce(\br).
\eea
In Eqs.~(\ref{SCPB1})-(\ref{mp0}), we introduced the polymer charge density $\sigma(\br)=-\sigma_{\rm p}\delta(r-a_*)$, the DH screening function associated with the weak-coupling (WC) salt,
\be
\label{dh}
\kappa(r)=\kappa\s e^{-V_i(\br)}\;;\hspace{5mm}\kappa\s^2=8\pi\ell_{\rm B}n_{+\rm b},
\ee
and the ion densities
\bea\label{salt}
n_{\pm}(\br)&=&n_{{\rm b}\pm}k_\pm(\br)\left[1+n_{\rm bc}T_\pm(\br)\right],\\
\label{mul}
n\ce(\br)&=&n_{\rm bc}k\ce(\br).
\eea
Eqs.~(\ref{salt})-(\ref{mul}) include the bare ionic partition functions
\be\label{part}
k_i(\br)=e^{-V_i(\br)-q_i\phi\s(\br)-q_i^2\delta G(\br)/2}
\ee
for $i=\{\pm,{\rm c}\}$, and the non-uniform virial coefficient
\be\label{vir}
T_i(\br)=\int\mathrm{d}^3\br\ce\left[k\ce(\br\ce)f_i(\br,\br\ce)-f_{i{\rm b}}(\br-\br\ce)\right]
\ee
taking into account the direct interactions between the WC salt and the SC multivalent charges, with the Mayer function and its bulk limit defined as
\bea
\label{may1}
f_i(\br,\br\ce)&=&e^{-w(\br-\br\ce)-q_iq\ce G(\br,\br\ce)}-1,\\
\label{may2}
f_{i{\rm b}}(\br-\br\ce)&=&e^{-w(\br-\br\ce)-q_iq\ce G_{\rm b}(\br-\br\ce)}-1.
\eea
In Eqs.~(\ref{dh}) and~(\ref{part}), the function $V_i(\br)$ is a steric potential imposing the impenetrability of the membrane and the polymer by the ions, and the function $w(\br-\br\ce)$ in Eqs.~(\ref{may1})-(\ref{may2}) is a general hard-core (HC) potential between the salt and multivalent ions.

Eq.~(\ref{part}) involves the self-energy defined as the equal point Green's function renormalized by its bulk value, 
\be\label{se}
\delta G(\br)=\left[G(\br,\br')-G\B(\br-\br')\right]_{\br'\to\br},
\ee
where the electrostatic Green's function satisfies the Debye-H\"{u}ckel-level kernel equation
\be
\label{gr1}
\left[\nabla\e(\br)\nabla-\e(\br)\kappa^2(r)\right]G(\br,\br')=-\frac{e^2}{k_{\rm B}T}\delta(\br-\br').
\ee
We also note that using the inverse Green's function
\be
G^{-1}(\br,\br')=-\frac{k_{\rm B} T}{e^2}\left[\nabla\e(\br)\nabla-\e(\br)\kappa^2(r)\right]\delta(\br-\br'),
\ee
Eq.~(\ref{mp0}) can be inverted to obtain the multivalent ion-induced potential as
\be
\label{mp}
\hspace{-3mm}\phi\ce(\br)=q\ce\int\mathrm{d}^3\br\ce n\ce(\br\ce)G(\br,\br\ce).
\ee
This implies that the calculation of the average potential~(\ref{sup}) requires the solution of the Green's Eq.~(\ref{gr1}). This task will be carried out in Appendix~\ref{sgr}.

One notes that in the bulk ion reservoir where the salt-dressed potential vanishes, $\phi\s(\br)\to0$, and the Green's function tends to its bulk value, i.e. $G(\br,\br')\to G\B(\br-\br')$, the local density functions~(\ref{salt})-(\ref{mul}) consistently converge to the reservoir concentration values, i.e. $n_i(\br)\to n_{i{\rm b}}$. Furthermore, we emphasize that in a spherically symmetric bulk liquid, the DH Eq.~(\ref{gr1}) is solved by the bulk Green's function
\be\label{grb1}
G\B(\br-\br')=\frac{4\pi\ell_{\rm B}}{|\br-\br'|}e^{-\kappa\s|\br-\br'|}
\ee
satisfying the equation $\int_{\br\ce}G\B(\br-\br\ce)=4\pi\ell_{\rm B}/\kappa\s^2$. Using the above equalities in Eqs.~(\ref{SCPB1}) and~(\ref{mp}), the bulk electroneutrality condition consistently follows as
\be
\label{be}
q_+n_{\rm b+}+q_-n_{\rm b-}+q\ce n_{\rm bc}=0.
\ee

We finally note that by combining Eqs.~(\ref{SCPB1})-(\ref{mp0}), one obtains a Poisson Eq. for total average potential~(\ref{sup}),
\be
\label{pois}
\frac{k_{\rm B}T}{e^2}\nabla\cdot\e(\br)\nabla\psi(\br)+\sum_{i=\pm,{\rm c}}q_in_i(\br)+\sigma(\br)=0.
\ee
In the main text of our article, the SCPB Eq.~(\ref{pois}) is used to derive the polymer and convective liquid velocities in the pore.

\section{Solving the Green's Eq.~(\ref{gr1})}
\label{sgr}

In this appendix, we solve Eq.~(\ref{gr1}) to obtain the electrostatic Green's function. In the cylindrical geometry of the nanopore confining the polyelectrolyte, the dielectric permittivity profile $\e(\br)$, and the steric ion potential $V_i(\br)$ in the screening function~(\ref{dh}) depend solely on the radial coordinate, i.e.
\bea
\label{ep}
&&\e(r)=\e_{\rm w}\theta(r-a)\theta(d-r)+\e_{\rm p}\theta(a-r)+\e\m\theta(r-d),\nonumber\\
&&\\
&&e^{-V_i(r)}=\theta(r-a_*)\theta(d_*-r),
\eea
with the effective polymer radius $a_*=a+l\s$ and nanopore radius $d_*=d-l\s$ renormalized by the length $l\s$. Hence, the Green's function can be Fourier-expanded as
\be
\label{gr2}
G(\br,\br')=\sum_{m=-\infty}^{+\infty}\int_{-\infty}^{+\infty}\frac{\mathrm{d}k}{4\pi^2}e^{im(\theta-\theta')}e^{ik(z-z')}\tG_m(r,r';k).
\ee

Injecting Eq.~(\ref{gr2}) into the Green's Eq.~(\ref{gr1}), the latter yields the Green's Eq. in Fourier basis,
\bea
\label{gr3}
&&\left\{\frac{1}{r}\partial_r\left[r\e(r)\partial_r\right]-\e(r)\left[\frac{m^2}{r^2}+p^2(r)\right]\right\}\tG_m(r,r';k)\nonumber\\
&&=-\frac{e^2}{k_{\rm B}Tr}\delta(r-r'),
\eea
where $p^2(r)=k^2+\kappa^2(r)$. Introducing the parameter $p=\sqrt{k^2+\kappa\s^2}$, the general solution of the differential Eq.~(\ref{gr3}) can be expressed in terms of the Bessel functions of the first kind $\mI(x)$ and second kind $\mK(x)$~\cite{math} as
\begin{widetext}
\bea
\label{gr4}
\tG_m(r,r';k)&=&c_1\mI(kr)\theta(a-r)+\left[c_2\mI(kr)+c_3\mK(kr)\right]\theta(r-a)\theta(a_*-r)\\
&&+\left[c_4\mI(pr)+c_5\mK(pr)\right]\theta(r-a_*)\theta(r'-r)+\left[c_6\mI(pr)+c_7\mK(pr)\right]\theta(r-r')\theta(d_*-r)\nonumber\\
&&+\left[c_8\mI(kr)+c_9\mK(kr)\right]\theta(r-d_*)\theta(d-r)+c_{10}\mK(kr)\theta(r-d),\nonumber
\eea
where the integration constants $c_i$ should be determined from the continuity of the Green's function $\tG_m(r,r';k)$ and the displacement field $\e(r)\partial_r\tG_m(r,r';k)$ at the points $r=\{a,a_*,d_*,d\}$, as well as the jump of the field at $r=r'$, i.e.
\be\label{gr5}
\left.\partial_r\tG_m(r,r';k)\right|_{r=r'_+}-\left.\partial_r\tG_m(r,r';k)\right|_{r=r'_-}=-\frac{4\pi\ell_{\rm B}}{r'}.
\ee
Introducing now the variables $r_<={\rm min}(r,r')$ and $r_>={\rm max}(r,r')$, the Green's function~(\ref{gr5}) reads for $a_*\leq r,r'\leq d_*$ 
\be\label{gr6}
\tG_m(r,r';k)=\tG_{m,{\rm b}}(r,r';k)+\delta\tG_m(r,r';k),
\ee
where the bulk component $\tG_{m{\rm b}}(r,r';k)=4\pi\ell_{\rm B}\mI(pr_<)\mK(pr_>)$ is the Fourier transform of the bulk DH potential~(\ref{grb1}), and the heterogeneous component accounting for the presence of the polymer and the pore reads
\be\label{gr7}
\delta\tG_m(r,r';k)=\frac{4\pi\ell_{\rm B}}{1-s_1s_2}\left\{s_1\mI(pr)\mI(pr')+s_2\mK(pr)\mK(pr')+s_1s_2\left[\mI(pr)\mK(pr')+\mI(pr')\mK(pr)\right]\right\}.
\ee
In Eq.~(\ref{gr7}), we introduced the auxiliary coefficients
\be\label{gr8}
s_1=\frac{\mK'(pd_*)-\gamma D\mK(pd_*)}{\gamma D\mI(pd_*)-\mI'(pd_*)};\hspace{1cm}s_2=\frac{\mI'(pa_*)-\gamma M\mI(pa_*)}{\gamma M\mK(pa_*)-\mK'(pa_*)},
\ee
with the parameters $\gamma=k/p$, $\eta_{\rm p}=\e_{\rm p}/\e_{\rm w}$, $\eta_{\rm m}=\e\m/\e_{\rm w}$, and
\bea
\label{gr9}
&&D=\frac{\mK(kd)\left[\mI'(kd_*)\mK'(kd)-\mK'(kd_*)\mI'(kd)\right]+\eta\m\mK'(kd)\left[\mI(kd)\mK'(kd_*)-\mI'(kd_*)\mK(kd)\right]}
{\mK(kd)\left[\mK'(kd)\mI(kd_*)-\mI'(kd)\mK(kd_*)\right]+\eta\m\mK'(kd)\left[\mI(kd)\mK(kd_*)-\mI(kd_*)\mK(kd)\right]},\\
&&M=\frac{\mI(ka)\left[\mI'(ka_*)\mK'(ka)-\mK'(ka_*)\mI'(ka)\right]+\eta_{\rm p}\mI'(ka)\left[\mI(ka)\mK'(ka_*)-\mI'(ka_*)\mK(ka)\right]}
{\mI(ka)\left[\mK'(ka)\mI(ka_*)-\mI'(ka)\mK(ka_*)\right]+\eta_{\rm p}\mI'(ka)\left[\mI(ka)\mK(ka_*)-\mI(ka_*)\mK(ka)\right]}.
\eea
\end{widetext}

Finally, injecting the Fourier expansion~(\ref{gr2}) into Eq.~(\ref{se}), the self-energy associated with the Fourier-transformed Green's function~(\ref{gr7}) becomes
\be
\label{se2}
\delta G(r)=\sum_{m=-\infty}^{+\infty}\int_{-\infty}^{+\infty}\frac{\mathrm{d}k}{4\pi^2}\delta\tG_m(r,r;k).
\ee

\section{Perturbative solution of the SCPB Eq.~(\ref{SCPB1})}

In the cylindrical nanopore geometry where the Green's function~(\ref{gr2}) corresponds to an infinite sum including an additional Fourier integral, the numerical evaluation of the virial function~(\ref{vir}) given by the three dimensional integral of the Mayer function~(\ref{may1}) is a formidable task. Therefore, the SCPB Eq.~(\ref{SCPB1}) is solved here via a perturbative approach. This perturbative scheme is based on the expansion of the SCPB Eq. in terms of the dilute multivalent ion concentration $n_{{\rm c}b}$. To this aim, we split the potential solving Eq.~(\ref{SCPB1}) into the pure salt-dressed component $\phi_0(\br)$ and the perturbative component $\phi_1(\br)$ of order $O\left(n_{\rm bc}\right)$ associated with the salt-multivalent ion screening,
\be
\label{vir1}
\phi\s(\br)=\phi_0(\br)+\phi_1(\br).
\ee
Then, we plug Eqs.~(\ref{gr2}) and~(\ref{vir1}) into Eqs.~(\ref{SCPB1})-(\ref{mul}), Taylor-expand the result at the linear order in the multivalent ion density $n_{{\rm c}b}$ and the potential component $\phi_1(\br)$, and use the electroneutrality condition~(\ref{be}) to eliminate the anion density. Finally, in Eqs.~(\ref{vir})-(\ref{may2}), we set the HC interaction potential to zero ($w(\br-\br\ce)\to0$), and carry out the Taylor expansion of the Mayer functions in terms of the Green's function to get $f_i(\br,\br\ce)\approx-q_iq\ce G(\br,\br\ce)$ and $f_{i{\rm b}}(\br-\br\ce)\approx-q_iq\ce G_{\rm b}(\br-\br\ce)$. Accounting for the cylindrical symmetry $\phi\s(\br)=\phi\s(r)$ associated with the purely radial surface charge density distribution $\sigma(r)=-\sigma_{\rm p}\delta(r-a_*)$, after lengthy algebra, the differential Eq. satisfied by the pure salt potential $\phi_0(r)$ follows at the order $O\left(n_{\rm bc}^0\right)$ as
\be
\label{vir3}
\frac{1}{r}\partial_rr\partial_r\phi_0(r)+\kappa_+^2\left[k_+(r)-k_-(r)\right]=4\pi\ell_{\rm B}\sigma_{\rm p}\delta(r-a_*),
\ee
and the Eq. solved by the perturbative potential component $\phi_1(r)$ of order $O\left(n_{\rm bc}\right)$ becomes
\be\label{vir4}
\frac{1}{r}\partial_rr\partial_r\phi_1(r)-\kappa_+^2\left[k_+(r)+k_-(r)\right]\phi_1(r)=-\kappa\ce^2J\ce(r).
\ee
In Eqs.~(\ref{vir3})-(\ref{vir4}), the auxiliary screening parameters and partition functions of the single ionic species read
\bea
\label{vir5}
&&\kappa_i^2=4\pi\ell_{\rm B}n_{{\rm b}i}q_i,\\
\label{vir6}
&&k_i(r)=e^{-V_i(r)-q_i\phi_0(r)-q_i^2\delta G(r)/2},
\eea
 for $i=\{\pm,{\rm c}\}$. Moreover, the source function is 
\bea
\label{vir7}
J\ce(r)&=&\frac{q_+}{q_+-q_-}\left[k_+(r)-k_-(r)\right]\\
&&-\frac{\kappa_+^2}{\kappa\ce^2}\left\{k_+(r)+k_-(r)-2\right\}e^{-V_i(r)}\phi\ce(r),\nonumber
\eea
with the multivalent ion-induced potential in Eq.~(\ref{mp})
\be
\label{vir8}
\phi\ce(r)=q\ce n_{\rm bc}\int_{a_*}^{d_*}\mathrm{d}r' r' k\ce(r')\tG_0(r,r';k=0).
\ee
Finally, upon the same perturbative expansion, the ion densities~(\ref{salt})-(\ref{mul}) follow as
\bea\label{vir9}
n_+(r)&=&n_{{\rm b}+}k_+(r)\left\{1-\left[\phi_1(r)+\phi\ce(r)-\frac{\kappa\ce^2}{\kappa\s^2}\right]\right\},\\
n_-(r)&=&\left\{n_{{\rm b}-}+n_{{\rm b}+}\left[\phi_1(r)+\phi\ce(r)-\frac{\kappa\ce^2}{\kappa\s^2}\right]\right\}k_-(r),\nonumber\\
&&\\
\label{vir7}
n\ce(r)&=&n_{\rm bc}k\ce(r).
\eea

Eq.~(\ref{vir3}) is equivalent to the modified PB Eq. augmented by the ionic image charge interactions. Integrating this equation around $r=a_*$ and $r=d_*$,  the boundary conditions to be included in the numerical solution scheme follow as
\be
\label{bou1}
\phi'\left(a_*^+\right)=4\pi\ell_{\rm B}\sigma_{\rm p}\;;\hspace{1cm}\phi'\left(d_*^-\right)=0.
\ee
We also note that for the numerical evaluation of Eq.~(\ref{vir8}), one needs the infrared limit of the Fourier-transformed Green's function~(\ref{gr6}) at $m=0$ and $k=0$ where the coefficients in Eqs.~(\ref{gr8}) reduce to $s_1={\rm K}_1(kd_*)/{\rm I}_1(kd_*)$ and $s_2={\rm I}_1(ka_*)/{\rm K}_1(ka_*)$.


\begin{thebibliography}{99}
\bibitem {gn1} Podgornik, R.; Strey, H. H.; Parsegian, V. A. \textit{Molecular Interactions in Lipids, DNA and DNA-lipid Complexes, in Gene Therapy: Therapeutic Mechanisms and Strategies}, 209-239, Marcel Dekker, New York, 2000.
\bibitem {gn2} Yin, H.; Kanasty, R.L.; Eltoukhy, A.; Vegas, A.J.; Dorkin, J.R.; Anderson, D.G. Non-viral vectors for gene-based therapy. {\it Nat. Rev. Genet.} {\bf 2014}, {\it 15}, 541–555.
\bibitem {gn3} Thomas, T.J.; Tajmirriahi, H.A. Polyamine-DNA interactions and development of gene delivery vehicles. {\it Amino Acids} {\bf 2016}, {\it 48}, 2423–2431. 
\bibitem{biomatter} Holm, C.; Kekicheff, P.; Podgornik, R. {\it Electrostatic Effects in Soft Matter and Biophysics}; Kluwer Academic, Dordrecht, 2001.
\bibitem {com}  Jain, M.; Olsen, H. E.; Paten, B.;  Akeson, M. The Oxford Nanopore MinION: delivery of nanopore sequencing to the genomics community. {\it Genome Biol.} {\bf 2016}, {\it 17}, 239.
\bibitem {rev2} Wanunu, W. Nanopores: A journey towards DNA sequencing. {\it Phys. Life Rev.} {\bf 2012}, {\it 9}, 125-158. 
\bibitem {Tapsarev} Palyulin, V. V.; Ala-Nissila, T.; Metzler, R. Polymer translocation: the first two decades and the recent diversification. {\it Soft Matter} {\bf 2014}, {\it 10}, 9016-9037.
\bibitem {e1}  Kasianowicz, J. J.;  Brandin, E.; Branton, D.; Deamer, D. W. Characterization of individual polynucleotide molecules using a membrane channel. {\it Proc. Natl. Acad. Sci. U.S.A} {\bf 1996}, {\it 93}, 13770-13773.  
\bibitem {e2} Meller, A.; Nivon, L.; Branton, D. Voltage-Driven DNA Translocations through a Nanopore. {\it Phys. Rev. Lett.} {\bf 2001}, {\it 86}, 3435-3438. 
\bibitem {e6}  Smeets, R. M. M.;  Keyser, U. F.; Krapf, D.; Wue, M.-Y.; Dekker, N. H.; Dekker, C. Salt dependence of ion transport and DNA translocation through solid-state nanopores. \textit{Nano Lett.} \textbf{2006}, \textit{6}, 89-95.
\bibitem {e3} Bonthuis, D. J.; Zhang, J.; Hornblower, B.; Math\'{e}, J.; Shklovskii, B. I.; Meller, A. Self-Energy-Limited Ion Transport in Subnanometer Channels. {\it Phys. Rev. Lett.} {\bf 2006}, {\it 97}, 128104. 
\bibitem {e7} Wanunu, M.; Sutin, J.; Mcnally, B.; Chow, A.; Meller, A. DNA Translocation Governed by Interactions with Solid-State Nanopores.  {\it Biophys. J.} {\bf 2008}, {\it 95}, 4716. 
\bibitem {e4} Clarke, J.;  Wu, H. C.; Jayasinghe, L.;  Patel, A.; Reid, S.; Bayley, H. Continuous base identification for single-molecule nanopore DNA sequencing. {\it Nature Nanotech.} {\bf 2009}, {\it 4}, 265-270. 
\bibitem {e5} Wanunu, M.; Morrison, W.; Rabin, Y.; Grosberg, A.Y. ; Meller, A. Electrostatic focusing of unlabelled DNA into nanoscale pores using a salt gradient. {\it Nature Nanotech.} {\bf 2010}, {\it 5}, 160-165. 
\bibitem {Coating2} Sischka, A.; Galla, L.; Meyer, A. J.; Spiering, A. Controlled translocation of DNA through nanopores in carbon nano-, silicon-nitride- and lipid-coated membranes. {\it Analyst} {\bf 2015}, {\it 140}, 4843-4847. 
\bibitem {Coating1} Eggenberger, O. M.; Ying, C.; Mayer, M. Surface coatings for solid-state nanopores. {\it Nanoscale} {\bf 2019}, {\it 11}, 19636. 
\bibitem {e8} Kang, J. H.; Lee, K.; Kim, H.M.;  Kim, K. B. Slowing DNA translocation through a solid- state nanopore by applying hydrophobic microchannel-guided walls. {\it J. Vac. Sci. Technol. A} {\bf 2020}, {\it 38}, 043401.
\bibitem {e10} Asandei, A.; Muccio, G.D.; Schiopu, I.; Mereuta, L.; Dragomir, I.S.; Chinappi, M.; Luchian, T. Nanopore-Based Protein Sequencing Using Biopores: Current Achievements and Open Challenges. {\it Small Methods} {\bf 2020}, {\it 1900595}.
\bibitem {e9} Yang, W.; Radha, B.; Choudhary, A.; You, Y.; Mettela, G.; Geim, A.K.;  Aksimentiev, A.; Keerthi, A.; Dekker, C. {\it Adv. Mater.} {\bf 2021}, 2007682.
\bibitem {Coating3} Wang, C.; Sensale, S.; Pan, Z.; Senapati, S.; Chang, H.-C. Slowing down DNA translocation through solid- state nanopores by edge-field leakage. {\it Nat. Commun.} {\bf 2021}, {\it 12}, 140. 
\bibitem {n1} Sung, W.; Park, P. J. Polymer Translocation through a Pore in a Membrane. {\it Phys. Rev. Lett.} {\bf 1996}, {\it 77}, 783. 
\bibitem {n2} Ikonen, T.; Bhattacharya, A.;  Ala-Nissila, T.; Sung, W. Unifying model of driven polymer translocation. {\it Phys. Rev. E} {\bf 2012}, {\it 85}, 051803. 
\bibitem {n5} Farahpour, F.; Maleknejad, A.; Varnikc, F.; Ejtehadi, M. R. Chain deformation in translocation phenomena. \textit{Soft Matter} \textbf{2013}, \textit{9}, 2750-2759. 
\bibitem {Wolterink} Wolterink, J.K.; Barkema, G.T.; Panja, D. Passage Times for Unbiased Polymer Translocation through a Narrow Pore. {\it Phys. Rev. Lett.} {\bf 2006}, {\it 96}, 208301.
\bibitem {Saka1} Sakaue, T. Nonequilibrium dynamics of polymer translocation and straightening. {\it Phys. Rev. E} {\bf 2007}, {\it 76}, 021803.
\bibitem {Saka2} Saito, T.; Sakaue, T. Dynamical diagram and scaling in polymer driven translocation. {\it Eur. Phys. J. E} {\bf 2011}, {\it 34} 135.
\bibitem {mut1} Wong, C.T.A.; Muthukumar, M.  Polymer capture by electro-osmotic flow of oppositely charged nanopores. {\it J. Chem. Phys.} {\bf 2007}, {\it 126}.
\bibitem {mut2} Muthukumar, M. Theory of capture rate in polymer translocation. {\it J. Chem. Phys.} {\bf 2010}, {\it 132}, 195101.
\bibitem {mut3} Bell, N. A. W.; Muthukumar, M.; Keyser, U.F. Translocation frequency of double-stranded DNA through a solid-state nanopore. {\it Phys. Rev. E} {\bf 2016}, {\it 93}, 022401.  
\bibitem {Chinappi2} Chinappi, M.; Luchian, T.; Cecconi, F. Nanopore tweezers: Voltage-controlled trapping and releasing of analytes. {\it Phys. Rev. E} {\bf 2015}, {\it 92}, 032714.
\bibitem {HLsim} Ansalone, P.; Chinappi, M.; Rondoni, L.; Cecconi, F. Driven diffusion against electrostatic or effective energy barrier across $\alpha$-hemolysin. \textit{J. Chem. Phys.} \textbf{2017}, \textit{143}, 154109.
\bibitem {cap1} Qiao, L.; Ignacio, M.; Slater, W. G. Voltage-driven translocation: Defining a capture radius. \textit{J. Chem. Phys.} \textbf{2019}, \textit{151}, 244902.
\bibitem {cap2} Chinappi, M.; Yamaji, M.; Kawano, R.; Cecconi, F. Analytical Model for Particle Capture in Nanopores Elucidates Competition among Electrophoresis, Electroosmosis, and
Dielectrophoresis. \textit{ACS Nano} \textbf{2020}, \textit{14}, 15816-15828.
\bibitem {aks1} Aksimentiev, A.; Heng, J.B.; Timp, G.; Schulten, K. Microscopic Kinetics of DNA Translocation through synthetic nanopores. \textit{Biophys. J.} {\bf 2004}, \textbf{87}, 2086-2097.
\bibitem {aks2} Luan, B.; Aksimentiev, A. Electro-osmotic screening of the DNA charge in a nanopore. \textit{Phys. Rev. E} {\bf 2008}, \textit{78}, 021912.  
\bibitem {aks3} Luan, B.; Aksimentiev, A. Electric and electrophoretic inversion of the DNA charge in multivalent electrolytes. \textit{Soft Matter} \textbf{2010}, \textit{6}, 243-246. 
\bibitem {Hsiao} Hsiao, P.Y.; Translocation of a Polyelectrolyte through a Nanopore in the Presence of Trivalent Counterions: A Comparison with the Cases in Monovalent and Divalent Salt Solutions. {\it ACS Omega} {\bf 2020}, {\it 5},19805-19819.
\bibitem {Ghosal2006} Ghosal, S. Electrophoresis of a polyelectrolyte through a nanopore. \textit{Phys. Rev. E} \textbf{2006}, \textit{74}, 041901.  
\bibitem {Ghosal2007} Ghosal, S. Effect of Salt Concentration on the Electrophoretic Speed of a Polyelectrolyte through a Nanopore. \textit{Phys. Rev. Lett.} \textbf{2007}, \textit{98}, 238104.
\bibitem {the16} Buyukdagli, S.; Ala-Nissila, T. Controlling Polymer Translocation and Ion Transport via Charge Correlations. \textit{Langmuir} \textbf{2014}, \textit{30}, 12907-12915.
\bibitem {Buyuk2018} Buyukdagli, S. Enhanced polymer capture speed and extended translocation time in pressure-solvation traps. {\em Phys. Rev. E} {\bf 2018}, {\em 97}, 062406.
\bibitem {Buyuk2018II} Buyukdagli, S. Facilitated Polymer Capture by Charge Inverted Electroosmotic Flow in Voltage-driven Polymer Translocation. {\it Soft Matter} {\bf 2018}, \textit{14}, 3541.
\bibitem {Firnkes} Firnkes, M.; Pedone, D.; Knezevic, J.; D\"{o}blinger, M.; Rant, U. Electrically facilitated translocations of proteins through silicon nitride nanopores: conjoint and competitive action of diffusion, electrophoresis, and electroosmosis. {\it Nano Lett. } {\bf 2010}, {\it 10}, 2162-2167.
\bibitem {Buyuk2018III} Buyukdagli, S.; Ala-Nissila, T. pH-mediated regulation of polymer transport through SiN pores. {\it Europhys. Lett.} {\bf 2018}. {\it 123}, 38003.
\bibitem {Coating4} Yuan, J.-K.; Yao, S.-H.; Dang, Z.-M.; Sylvestre, A.; Genestoux, M.; Bai, J. Giant Dielectric Permittivity Nanocomposites: Realizing True Potential of Pristine Carbon Nanotubes in Polyvinylidene Fluoride Matrix through an Enhanced Interfacial Interaction. {\it J. Phys. Chem. C} {\bf 2011}, \textit{115}, 5515. 
\bibitem {Coating5} Chang, J.; Liang, G.; Gu, A.; Cai, S.; Yuan, L. The production of carbon nanotube/epoxy composites with a very high dielectric constant and low dielectric loss by microwave curing. {\it Carbon} {\bf 2012}, \textit{50}, 689-698. 
\bibitem {NetzSC} Moreira, A.G.; Netz, R.R. Strong-coupling Theory for Counter-ion Distributions. {\it Europhys. Lett.} {\bf 2000}, \textit{52}, 705. 
\bibitem {Podgornik2010} Kandu\v c, M.; Naji, A.; Forsman, J.; Podgornik, R. Dressed Counterions: Strong Electrostatic Coupling in the Presence of Salt. {\it J. Chem. Phys.} {\bf 2010} \textit{132}, 124701. 
\bibitem {Buyuk2019} Buyukdagli, S.; Podgornik, R. Like-charge Polymer-membrane Complexation Mediated by Multivalent Cations: One-loop-dressed Strong Coupling Theory. {\it J. Chem. Phys.} {\bf 2019}, \textit{151}, 094902. 
\bibitem {JPCB2020} Buyukdagli, S. Nanofluidic Charge Transport under Strong Electrostatic Coupling Conditions. {\it J. Phys. Chem. B} {\bf 2020}, \textit{124}, 11299-11309.
\bibitem {Boc1} Barrat, J.-L.; Bocquet, L. Large Slip Effect at a Nonwetting Fluid-Solid Interface. \textit{Phys. Rev. Lett.}, {\bf 1999}, {\it 82}, 4671-4674.
\bibitem {Boc2} Sendner, C.; Horinek, D.; Bocquet, L.; Netz, R. Interfacial Water at Hydrophobic and Hydrophilic Surfaces: Slip, Viscosity, and Diffusion. \textit{Langmuir} {\bf 2009}, {\it 25}, 10768-10781.
\bibitem {Heyden2006} van der Heyden, F. H. J.; Bonthuis, D. J.; Stein, D.; Christine, M.; Dekker, C. Electrokinetic energy conversion efficiency in nanofluidic channels. {\it Nano Lett.} {\bf 2006}, {\it 6}, 2232. 
\bibitem {Gillespie2013} Hoffmann, J.; Gillespie, D. Ion Correlations in Nanofluidic Channels: Effects of Ion Size, Valence, and Concentration on Voltage-and Pressure-driven Currents. {\it Langmuir} {\bf 2013}, {\it 29}, 1303. 
\bibitem {nt1} Ion transport experiments indicate that the surface-coating of membrane nanopores substantially suppresses the effect of the fixed surface charges~\cite{Coating1}. Therefore, in order to simplify the analysis, we consider here interfacially neutral nanopores.
\bibitem {ionm} The mobilities of the ion species \BS{${\rm C}^+$} and \BS{$\rm{A}^-$} are fixed to the experimentally measured values of $\mu_+=7.616\times10^{-8}$ $\mbox{m}^2\mbox{V}^{-1}\mbox{s}^{-1}$ and  $\mu_-=7.909\times10^{-8}$ $\mbox{m}^2\mbox{V}^{-1}\mbox{s}^{-1}$ for ${\rm K}^+$ and $\rm{Cl}^-$, respectively~\cite{book}.  As the drift mobilities of \BS{trivalent cations such as} spermidine ${\rm Spm}^{3+}$ \BS{or cobalt $\rm{Co}^{3+}$} are not available in the literature, the transport coefficient of the trivalent cation species \BS{${\rm M}^{3+}$} was estimated via the Einstein relation $\mu_i=|q_i|D/(k_{\rm B}T)$ as $\mu_{3+}=3\mu_+$.
\bibitem {book}  Lide, D. R. \textit{Handbook of Chemistry and Physics}, 93th edition, \textbf{2012}, CRC Press.
\bibitem {exp2} Hoogerheide, D. P.; Lu, B.; Golovchenko, J. A. Pressure-Voltage Trap for DNA near a Solid-State Nanopore. {\it ACS Nano} {\bf 2014}, {\it 8}, 7384 (2014).
\bibitem {nt3} It is noteworthy that a similar multivalent ion-driven monovalent charge discrimination has been previously observed in the MD simulations of electrolyte mixtures confined to cylindrical nanopores~\cite{Boda2007}. 
\bibitem {Boda2007} Boda, D.; Valisko, M.; Eisenberg, B.; Nonner, W.; Henderson, D.; Gillespie, D. Combined Effect of Pore Radius and Protein Dielectric Coefficient on the Selectivity of a Calcium Channel. {\it Phys. Rev. Lett.} {\bf 2007}, \textit{98}, 168102.
\BS{\bibitem {ExpHey} van der Heyden, F. H. J.; Stein, D.; Besteman, K.; Lemay, S. G.; Dekker, C. Charge Inversion at High Ionic Strength Studied by Streaming Currents. \textit{Phys. Rev. Lett.} \textbf{2006}, \textit{96}, 224502.}   
\bibitem {poltrex1} Qiu, S.; Wang, Y.; Cao, B.; Guo, Z.;  Chen, Y.; Yang, G. The suppression and promotion of DNA charge inversion by mixing counterions. {\it Soft Matter} {\bf 2015}, {\it 11}, 4099-4105.  
\BS{\bibitem {poltrex2} Wang, Y.; Wang, R.; Cao, B.; Guo, Z.; Yang, G. Single Molecular Demonstration of Modulating Charge Inversion of DNA. {\it Scientific Reports} {\bf 2016}, {\it 6}, 38628.}
\bibitem {poltrex3} Wang, Y.; Wang, R.; Gao, T.; Yang, G. The Mixing Counterion Effect on DNA Compaction and Charge Neutralization at Low Ionic Strength. {\it Polymers} {\bf 2018}, \textit{10}, 244. 
\bibitem {nt2} In Eq.~(\ref{cumch}), the second equality follows from the integration of the SCPB  \SB{equation}~(\ref{m1}).
\BS{\bibitem {pers} Brunet, A.; Tardin, C.; Salome, L.; Rousseau, P.; Destainville, N.; Manghi, M. Dependence of DNA Persistence Length on Ionic Strength of Solutions with Monovalent and Divalent Salts: A Joint Theory-Experiment Study. {\it Macromolecules} {\bf 2015}, {\it 48}, 3641-3652.}
\BS{\bibitem {nonloc} Buyukdagli, S.; Ala-Nissila, T. Microscopic formulation of non-local electrostatics in polar liquids embedding polarizable ions. {\it Phys. Rev. E} {\bf 2013}, {\it 87}, 063201.}
\BS{\bibitem {DipFix} Buyukdagli S.; Podgornik, R. Contribution of dipolar bridging to phospholipid membrane interactions: A mean-field analysis. {\it Phys. Rev. E} {\bf 2020}, {\it 102}, 012806.}
\end{thebibliography}
\end{document}